\documentclass[sigconf]{acmart}
\AtBeginDocument{%
  \providecommand\BibTeX{{%
    \normalfont B\kern-0.5em{\scshape i\kern-0.25em b}\kern-0.8em\TeX}}}
\usepackage{dblfloatfix}
\usepackage[bottom]{footmisc}
\usepackage{enumitem}
\setlist[itemize]{leftmargin=*, topsep=3pt}
\usepackage{xcolor} 
\usepackage{makecell}
\usepackage{subcaption}
\usepackage{xspace}
\usepackage[inkscapelatex=false]{svg}
\usepackage{soul}
\usepackage{listings}
\definecolor{codegreen}{rgb}{0,0.6,0}
\definecolor{codegray}{rgb}{0.5,0.5,0.5}
\definecolor{codepurple}{rgb}{0.58,0,0.82}
\definecolor{backcolour}{rgb}{0.95,0.95,0.95}

\lstdefinestyle{mystyle}{
    backgroundcolor=\color{backcolour},   
    commentstyle=\color{codepurple},
    keywordstyle=\color{NavyBlue},
    numberstyle=\tiny\color{codegray},
    stringstyle=\color{codepurple},
    basicstyle=\ttfamily\tiny,
    breakatwhitespace=true,         
    breaklines=true,                 
    captionpos=t,                    
    keepspaces=false,                 
    numbers=left,                    
    numbersep=5pt,                  
    showspaces=false,                
    showstringspaces=false,
    showtabs=false,                  
    tabsize=2
}

\newcommand{\alphaval}[2]{{\small $p\,#1\,#2$}}

\newcommand{\friedman}[4]{{\small [$\chi^{2}(#1)=#2$, \alphaval{#3}{#4}}]}
\newcommand{\wilcoxon}[3]{{\small [$Z=#1$, \alphaval{#2}{#3}}]}
\newcommand{\ttest}[4]{{\small [$t(#1)=#2$, \alphaval{#3}{#4}]}}

\newcommand{\ptps}[1]{\textsc{ptps#1}}
\newcommand{\llms}[1]{\textsc{vlms#1}}

\newcommand{\ivmethod}[1]{\textsc{method type#1}}
\newcommand{\ivscenario}[1]{\textsc{scenario#1}}
\newcommand{\cours}[1]{\emph{AutoOptimization#1}}
\newcommand{\cbaseline}[1]{\emph{ParetoAdapt#1}}
\newcommand{\cpareto}[1]{\emph{ParetoAdapt#1}}
\newcommand{\cmanual}[1]{\emph{ManualPlace#1}}

\newcommand{\anova}[6]{{\small [$F_{#1,#2}$\,$=$\,$#3$, $p$\,$#4$\,$#5$]}}
\newcommand{\pvall}[2]{{\small $p\,#1\,#2$}} 

\copyrightyear{2026}
\acmYear{2026}
\setcopyright{cc}
\setcctype{by-nc-nd}
\acmConference[CHI '26]{Proceedings of the 2026 CHI Conference on Human Factors in Computing Systems}{April 13--17, 2026}{Barcelona, Spain}
\acmBooktitle{Proceedings of the 2026 CHI Conference on Human Factors in Computing Systems (CHI '26), April 13--17, 2026, Barcelona, Spain}
\acmPrice{}
\acmDOI{10.1145/3772318.3791444}
\acmISBN{979-8-4007-2278-3/2026/04}

\newcommand{\projname}{AutoOptimization\xspace}
\newcommand{\projectname}{\projname}
\newcommand\sscnt[1]{\thesubsection.{#1}\hspace{1.2ex}}
\begin{document}

\title[Automating UI Optimization through Multi-Agentic Reasoning]{Automating UI Optimization through Multi-Agentic Reasoning}

\author{Zhipeng Li}
\affiliation{%
  \institution{Department of Computer Science}
  \institution{ETH Z\"urich}
  \country{Z\"urich, Switzerland}
}
\email{zhipeng.li@inf.ethz.ch}

\author{Christoph Gebhardt}
\affiliation{%
  \institution{Department of Computer Science}
  \institution{ETH Z\"urich}
  \country{Z\"urich, Switzerland}
}
\email{christoph.gebhardt@inf.ethz.ch}

\author{Yi-Chi Liao}
\affiliation{%
  \institution{Department of Computer Science}
  \institution{ETH Z\"urich}
  \country{Z\"urich, Switzerland}
}
\email{yichi.liao@inf.ethz.ch}

\author{Christian Holz}
\affiliation{%
  \institution{Department of Computer Science}
  \institution{ETH Z\"urich}
  \country{Z\"urich, Switzerland}
}
\email{christian.holz@inf.ethz.ch}

\begin{teaserfigure}
  \vspace{-2.5mm}%
  \includegraphics[width=\textwidth]{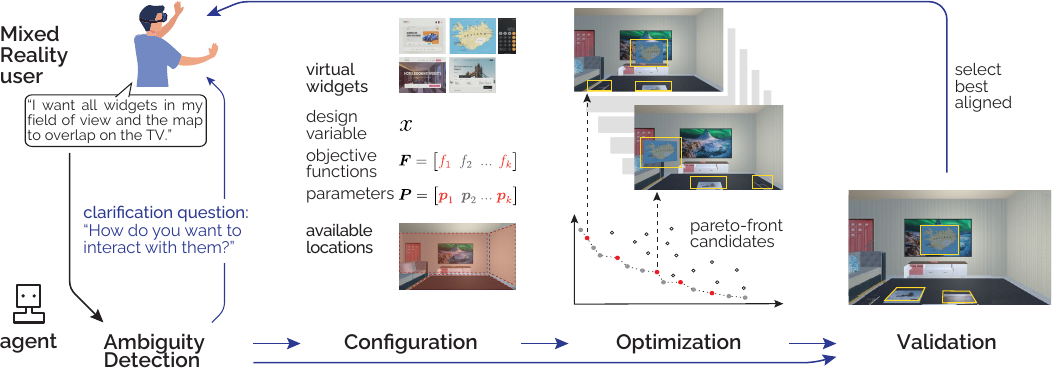}%
  \vspace{-2.5mm}%
  \caption{\projectname is a framework for \emph{personalizing optimization-based UI adaptations}. 
    From the user's stated instructions, we infer relevant terms and parameters to define a multi-objective optimization.
    First, \emph{Ambiguity Detection} clarifies instructions with the user as needed, such that \emph{Configuration} can specify the optimization problem by selecting the relevant objectives, UI elements, and placement locations. 
    For \emph{Optimization}, we run a solver to generate a set of Pareto-optimal layout candidates, and \emph{Validation} selects the one layout from the set that best aligns with the user's instructions.
  }%
  \Description{A flowchart depicting the process of widget placement optimization in a virtual reality (VR) environment. The flowchart includes five stages: Ambiguity Detection, Configuration, Optimization, and Validation. The first stage shows a user with a VR headset pointing at a screen displaying various widgets and asking “How do you want to interact with the item?” Below this are two boxes labeled “classification question” and “objective forms.” The second stage shows multiple widget options on different screens. The third stage has two screens; one labeled “not available locations” and another showing selected widgets in red outlines indicating selected locations. The fourth stage displays an algorithmic process for optimizing widget placement. Finally, the fifth stage shows three screens with progressively fewer widgets being validated for final placement in the VR environment.}
  \label{fig:teaser}
  \Description{A flowchart depicting the process of widget placement optimization in a virtual reality (VR) environment. The flowchart includes five stages: Ambiguity Detection, Configuration, Optimization, and Validation. The first stage shows a user with a VR headset pointing at a screen displaying various widgets and asking “How do you want to interact with the item?” Below this are two boxes labeled “classification question” and “objective forms.” The second stage shows multiple widget options on different screens. The third stage has two screens; one labeled “not available locations” and another showing selected widgets in red outlines indicating selected locations. The fourth stage displays an algorithmic process for optimizing widget placement. Finally, the fifth stage shows three screens with progressively fewer widgets being validated for final placement in the VR environment.
}
\end{teaserfigure}
\renewcommand{\shortauthors}{Li, et al.}
\begin{abstract}
    We present \emph{\projname}, a novel multi-objective optimization framework for adapting user interfaces.
From a user’s verbal preferences for changing a UI, our framework guides a prioritization-based Pareto frontier search over candidate layouts.
It selects suitable objective functions for UI placement while simultaneously parameterizing them according to the user's instructions to define the optimization problem.
A solver then generates a series of optimal UI layouts, which our framework validates against the user's instructions to adapt the UI with the final solution.
Our approach thus overcomes the previous need for manual inspection of  layouts and the use of population averages for objective parameters.
We integrate multiple agents sequentially within our framework, enabling the system to leverage their reasoning capabilities to interpret user preferences, configure the optimization problem, and validate optimization outcomes.
We evaluate each step of our framework inside a Mixed Reality use case and demonstrate that \emph{\projname} effectively increases the usability of UI adaptation schemes.
\vspace*{-2mm}

\end{abstract}





\newcommand{\figurewidth}{\linewidth} 
\newcommand{\change}[1]{\textcolor{black}{#1}}

\maketitle
\section{Introduction}
\label{sec:introduction}

Optimization has become a common approach for user interface (UI) adaptation.
It operates by defining objective functions that specify what constitutes a good interface for a given application scenario, taking into account factors such as users' context or ergonomic considerations. 
A solver then adapts the UI, i.e., repositioning, showing, or hiding UI elements, to meet these objectives.
Compared to manual exploration and adjustment, optimization has demonstrated a scalable and efficient solution for many UI adaptation problems, especially in Virtual Reality (VR) and Mixed Reality (MR) settings, where the interface design space is large~\cite{semanticadapt2021cheng,cheng2023interactionadapt,ens2015spatial,evangelista2022auit}.
Recently, multi-objective optimization (MOO) has been introduced to tackle UI layout adaptation problems. 
Rather than identifying a single best solution, MOO generates a set of Pareto-optimal designs, each representing a unique balance between competing objectives such as visibility, accessibility, and contextual relevance~\cite{paretooptimal2023johns}. 
These methods typically define fixed objective functions that incorporate scene semantics, physical affordances, or spatial consistency~\cite{semanticadapt2021cheng,cheng2023interactionadapt,ens2015spatial}.
This enhancement broadens the applicability of optimization methods by supporting scenarios where multiple goals must be balanced.

Despite the advantages, conducting optimization remains a labor-intensive and complex process for both designers (or developers) and users.
Designers must pre-define the design parameters (i.e., decision variables to be optimized) and the objective functions (i.e., metrics for evaluating design quality), all tailored to the scene and context. 
The objective terms and their parameters (such as usage frequency~\cite{li2024situationadapt}, arm length~\cite{cheng2023interactionadapt}, or semantic association penalties~\cite{semanticadapt2021cheng}) are often hardcoded or based on population-level averages. 
Therefore, current optimization approaches cannot adapt to dynamic user needs or task-specific contexts, and instead rely on designers to configure them for each scenario.

For users, validating the final output can be equally demanding. 
When MOO is used, the result is often a collection of Pareto-optimal designs, each of which is a potential ``best'' design under a different trade-off. The user must manually evaluate and compare these options to determine which aligns best with their preferences ~\cite{paretooptimal2023johns}.
While the weighted-sum approach can minimize the user's effort in exhaustively validating the Pareto-optimal designs ~\cite{evangelista2022auit}, this approach relies on the assigned weights and potentially leads to a sub-optimal outcome. 
These steps show that optimization is not truly automatic. 
In fact, it often relies heavily on the designer's expertise (to set up the optimization task) and the user's effort (to explore and select from the output). 
Because of these manual burdens, optimization workflows cannot operate dynamically or autonomously across changing contexts without human supervision and reconfiguration.
This brings us to a fundamental question:
Can we \emph{automate the optimization process to minimize the cumbersome aspects of UI adaptation for both designers and users}, and ultimately \emph{enable a more dynamic, streamlined experience of adaptive UIs}?


In this paper, we introduce \emph{\projname}, an optimization framework that leverages modern vision-language models (VLMs) as agents to streamline the cumbersome tasks of parameterizing and evaluating candidate designs in UI optimization.
In this framework, the VLM plays the role of a \emph{proxy designer}: First, it interprets the user’s needs and the surrounding context, asking follow-up questions when clarification is needed, and then automatically configures the optimization task.
Second, when the solver produces a set of Pareto-optimal designs, the VLM evaluates these options and selects the final design based on its understanding of the user's intent.
\change{We call this framework ``\textit{\projname}'' (short for Automating Optimization), as it is, to our knowledge, the first attempt to fully automate both the setup and decision-making phases of optimization through sequentially operating agents, reducing the user’s involvement to the convenient act of supplying context verbally.}
Our goal is to shift the optimization workflow from being \emph{human-driven} to \emph{foundation model-assisted}, significantly reducing humans' effort in the process.

Our framework consists of two computational components: a vision-language model (VLM) and an optimization module.
Both work in tandem to drive the UI adaptation process. 
\change{The VLM agents work in a sequential, pipeline-based manner to interpret the user's verbal instructions, reason about the contextual intent, and make high-level design decisions.}
The optimization module then computationally generates candidate UI configurations. 
Finally, the VLM evaluates these results and selects the most suitable design.

Figure~\ref{fig:teaser} illustrates the full process, starting with the user providing a high-level instruction or contextual goal, e.g., ``I want all widgets in my field of view and the map to overlap on the TV.'' 
It then proceeds through four sequential steps:
The first stage is \textit{ambiguity detection}, where the VLM identifies potential ambiguities in the user input and seeks clarification through follow-up questions until it gathers enough information to configure the optimization task. 
Once the context is clarified, the VLM proceeds to \textit{task initialization}, determining the relevant UI elements, design parameters, and objective functions based on the scene and user intent. 
The \textit{optimization module} then solves the problem, producing a set of Pareto-optimal designs that represent trade-offs among multiple objectives. 
In the final step, the VLM \textit{validates} these results and selects the most suitable design according to the original user-provided goals, effectively completing the end-to-end optimization process with minimal human intervention.

To evaluate the effectiveness of our \textit{\projname} framework in a representative use case, we demonstrate its application to MR UI layout adaptation.
To this end, we evaluated the individual stages of our framework. 
First, we collected data from experienced MR users on how they would instruct an intelligent assistant in various scenarios, categorizing their responses into well-informed or ambiguous instructions.
By learning from the collected responses, the VLM can detect the ambiguity in instructions with an accuracy 91.26\% in leave-one-user-out cross-validation and an accuracy of 92.93\% in leave-one-scenario-out cross-validation.
Next, we evaluated the performance of the VLM in selecting the layout that best aligns with user instructions from a set of Pareto-optimal candidates.
We collected decisions from 26 participants and 26 instances of the VLM, evaluating 72 candidates across 18 scenarios. 
The results indicate that the VLM instances identified the most appropriate UI layouts similarly to the participants' selections.
Finally, we evaluated our framework with an end-to-end user study.
Participants customized MR UI layouts using \projname, a state-of-the-art baseline~\cite{johns2023towards}, and a purely manual placement method.
The layouts generated by \projname demonstrated a closer alignment with user preferences, requiring fewer and smaller adjustments compared to the baseline methods.
Moreover, \projname achieved a comparable user satisfaction to manual placement while significantly reducing the effort. 
To summarize, we make these contributions in this paper:
\begin{itemize}[leftmargin=*,topsep=3pt]
    \item \textit{\projname}, a novel framework that shifts UI adaptation from human-driven to foundation model-assisted workflows. 
    Specifically, VLMs act as proxy-designers to automate key steps of an optimization process: input ambiguity detection, problem initialization, and result validation.
    \item an implementation of \projname for the use case of adaptive MR layout design. We also demonstrate its effectiveness and advantages over baseline approaches in a user study.
\end{itemize}

\section{Related work}

\projname is related to adaptive UIs in MR, their customization, and LLM use in HCI and problem solving.

\subsection{Adaptive User Interfaces in Mixed Reality}


Previous work has explored MR UI adaptation to the user’s state~\cite{visualattentionoptimization2019alghofaili,lu2022exploring,fender2017heatspace,veras2021elbow}, for example by adjusting visibility, placement, or detail level based on cognitive load~\cite{lindlbauer2019context} or gaze activity~\cite{gazeRL2019Gebhardt}. Other work has optimized MR interfaces ergonomically using quantitative pose recommendations~\cite{xrgonomics2021belo,ergo2017maurillo}.

Much MR interface optimization has focused on environment-based adaptation, aligning content with surroundings~\cite{cheng2022comfortable}, either using geometry~\cite{hettiarachchi2016annexing,jones2014roomalive} or specific objects like windows and walls~\cite{lages2019walking}. SemanticAdapt adapted interfaces based on semantic relationships between virtual and physical objects~\cite{semanticadapt2021cheng}, while InteractionAdapt considered input modes, positioning UI elements to leverage nearby surfaces for passive haptic feedback and facilitate cursor access~\cite{cheng2023interactionadapt}. 
\citeauthor{langerak2024marlui} proposed a multi-agent reinforcement learning framework that dynamically adapts point-and-click user interfaces by coordinating specialized agents to optimize layout and interaction efficiency in real time~\cite{langerak2024marlui}.
More recently, MineXR was proposed to support researchers in designing MR interfaces across various contexts~\cite{cho2024minexr}.

Although prior approaches effectively adapted layouts to the physical environment, they did not support tailoring UIs to individual users’ needs and preferences, assuming instead that objectives and their parameters are fixed a priori across users and contexts.
Because this assumption may not hold in real-world applications, our framework introduces the explicit capability of customizing the UI adaptation process based on the user's instructions, selecting a subset of objective terms and defining its parameters for optimal UI adaptation.



\subsection{Customization of Adaptive User Interfaces}
\label{sec:personalization-auis}

Adaptive MR systems often frame UI adaptation as a multi-objective problem, where static objective functions govern the placement of UI elements in the 3D environment. Since these functions apply uniformly to all users, prior research has focused on two approaches for accommodating user preferences: 
1)~adjusting objective weights and 2)~exploring the Pareto frontier for optimal solutions.

\emph{1)~A weighted sum of objective terms} let users express the relative importance of each objective via weight adjustment. 
By manually setting weights, the multi-objective problem is reduced to a single-objective one, yielding a single solution. 
For example, MenuOptimizer’s Objective Space Panel allows designers to set weights for objectives like performance, consistency, and similarity~\cite{bailly2013menuoptimizer}, while OPTIMISM’s GUI lets designers and end-users specify objective and heuristic weights for fabrication tasks~\cite{hofmann2023optimism}. 
In online 3D UI adaptation, pipelines such as SemanticAdapt~\cite{semanticadapt2021cheng}, InteractionAdapt~\cite{cheng2023interactionadapt}, and AUIT~\cite{evangelista2022auit} provide Unity control panels for adjusting individual objective weights. 
Although this manual, trial-and-error process is straightforward, collapsing multiple objectives into one can obscure Pareto-optimal solutions.

\emph{2)~Pareto-optimal methods}, in contrast, generate the full set of solutions balancing potentially competing objectives, allowing users to explore and select preferred designs. 
For instance, ParetoAdapt visualizes the Pareto frontier and lets users choose designs that best match their preferences~\cite{johns2023towards}. 
To reduce the manual effort of exploring trade-offs, our previous work proposed inferring user preferences from minimal UI element adjustments~\cite{yao2025preferenceguided} and translated them into priority levels for a priority-based MOO algorithm that automatically selects the final design from the Pareto frontier.
Other strategies include clustering to reduce solution sets, such as grouping haptic device designs by information transfer and recognition accuracy~\cite{liao2023interaction} or clustering input techniques by performance metrics~\cite{liao2021computational}. 
Both approaches enable users to prioritize objectives before exploring solutions.

Unlike previous approaches, \projname eliminates the need for users to manually select from a set of designs. 
Instead, 
our framework builds on a dynamic formulation of the underlying MOO problem and guides candidate layout selection via the user's instructions.
This way, \projname creates layouts that are tailored to the individual needs and preferences of users.

\subsection{LLMs in HCI \& Problem Solving}

LLMs have been widely used in HCI research to simulate users and solve user-centric problems.
In this paper, we leverage LLMs to reason about and simulate users’ preferences to automatically solve UI optimization problems.

Before simulating preferences, LLMs must first gather signals to understand users’ intentions. 
However, verbal instructions are often ambiguous. 
Therefore, LLMs need to detect and resolve ambiguities. 
\citeauthor{wang2023enabling} demonstrated that LLMs can ask users for missing input and answer UI-specific questions without task-specific training, showing their capability to understand user instructions and UI layouts~\cite{wang2023enabling}. 
AmbigChat further decomposes ambiguous questions into a hierarchical “disambiguation tree,” surfaces interactive UI widgets for each facet, and guides users through a conversational interface to clarify intentions~\cite{ma2025ambigchat}.

Beyond ambiguity detection, LLMs can simulate user preferences. 
Aligning LLMs to human preferences has been explored in several contexts.
\citeauthor{hamalainen2023evaluating} prompted GPT to provide open-ended responses about video game experiences, finding results comparable to human participants~\cite{hamalainen2023evaluating}.
\citeauthor{schmidt2024simulating} noted that LLM-generated survey responses, while sometimes artificial, reveal insights into human preferences and survey design flaws~\cite{schmidt2024simulating}. 
\citeauthor{kang2023llms} showed that LLMs, when fine-tuned on small interaction datasets, can reach or slightly surpass traditional recommendation models on predicting user ratings~\cite{kang2023llms}. 
Similarly, \citeauthor{liu2024crowdgenui} integrated a crowdsourced library of user preferences into LLM reasoning and code generation, enabling automated UI widgets that align better with real user intentions compared to LLM-only methods~\cite{liu2024crowdgenui}.

LLMs are also used to tackle complex UI tasks. 
As one of the most classic methods, Chain-of-Thought prompting allows intermediate reasoning steps to guide problem-solving~\cite{wei2022chain}, and majority-voting across multiple chains further improves outcomes~\cite{wang2022self}. 
Following these, Tree-of-Thoughts extends LLMs' problem solving capability by structuring reasoning as a tree with self-evaluation, enabling backtracking and strategic decision-making~\cite{yao2024tree}. 
As applications of these LLM-based reasoning frameworks, several studies investigate how to apply them to UI tasks. 
Guardian~\cite{ran2024guardian} uses LLMs to automate UI testing by prompting the model to plan each step needed to complete a task and then executing those steps to evaluate the interface. VisionTasker~\cite{song2024visiontasker} uses a VLM to convert UI screenshots into natural-language descriptions, and then plan the next action, enabling end-to-end automated task execution.

In optimization-based UI adaptation systems, LLMs can be leveraged to directly tackle the optimization problem itself. 
\citeauthor{yang2023large} demonstrated that LLMs can replace traditional optimizers by proposing candidate solutions in natural language and iteratively refining them, achieving competitive performance on small-scale optimization tasks~\cite{yang2023large}. 
Instead of fully replacing optimizers, another approach is to combine LLMs with existing optimization methods. 
For example, EvoPrompt integrates evolutionary algorithms with LLMs to optimize discrete prompts, leveraging both the search efficiency of evolutionary algorithms and the language coherence of LLMs~\cite{guo2023connecting}. 
Additionally, LLMs can provide parameters or contextual guidance for optimization tasks. 
For instance, SituationAdapt used a vision-language model to simulate user evaluations of virtual UI layouts~\cite{li2024situationadapt}, though it still relies on fixed weighted-sum objective functions, limiting its ability to adapt to dynamic user preferences.

Building on these prior works, our framework leverages a Chain-of-Thought approach to translate expressed user preferences into the formal context of a multi-objective optimization problem, automating the optimization-based UI adaptation process.

\section{\scalebox{1.1}{A}\scalebox{0.9}{uto}\scalebox{1.1}{O}\scalebox{0.9}{ptimization} framework}
\label{sec:framework}

Our framework tightly integrates user-specific subjective instructions with a multi-objective optimization problem by means of an VLM-based agent to minimize human involvement.
\autoref{fig:teaser} shows an overview of our framework, in which multiple agents are tasked with high-level decision-making to guide four phases:

For \emph{Ambiguity Detection}, a first agent processes user instructions to extract essential information and preferences for adapting the UI; if it determines the user-provided input to be insufficient, it iteratively prompts the user for clarification and complete instructions.
During \emph{Configuration}, the second agent specifies the optimization problem, including involved UI widgets, prioritized objectives, and the corresponding parameter values. 
Our agent then runs an \emph{Optimization} solver to search the Pareto front of the specified problem and generates multiple Pareto-optimal layout candidates.
For \emph{Validation}, a third agent validates these optimal layouts against the user's original instructions and selects the most suitable layout.
Below, we explain each of these components in more detail.

\subsection{Ambiguity Detection: Clarifying user instructions for completeness}
\label{sec:pipeline_ambiguity}

The input to our framework consists of an instruction through which a user describes how the UI should be structured, presented, or updated.
We leverage verbal instructions, because humans are used to expressing their preferences in this manner, while VLMs can process such natural language as input for reasoning tasks.

When a user provides an instruction, the Amiguity Detection agent evaluates whether the input is sufficiently clear and complete to infer layout preferences. 
To do so, the agent is also provided with the current UI layout, screenshots and names of all available widgets, and a screenshot of the user’s field of view. 
This visual context enables the agent to ground the user’s instruction in the actual interface state and to identify if essential details remain under-specified, including \textbf{which widgets are involved}, \textbf{how the user intends to interact with them}, and \textbf{user preferences}. 
The agent operates in an iterative refinement loop: if any ambiguity is detected, it asks the user a targeted clarification question, appends the response to the existing instruction, and re-evaluates the combined input. 
Through this accumulating context, the agent maintains a coherent understanding of the evolving instruction and continues the cycle until all relevant aspects are unambiguous, at which point it proceeds to the layout optimization stage.


The purpose of this dialog is to resolve ambiguity before proceeding to the actual optimization task.
Our agent is equipped with ten clarification questions as an example for the VLM to respond to the user for resolving ambiguity in the original instruction. 
We adapted these questions from prior research~\cite{kim2024aligning, mehrabi2023resolving} and tailored them to the task of UI adaptation.






\subsection{Configuration: Specifying the multi-objective optimization problem}

With the complete set of user instructions to inform the layout adaptation task, our framework's second component translates them into a multi-objective optimization problem.
For this purpose, a second agent dynamically includes or excludes individual terms from a set of optimization functions as well as UI elements, physical placement options, and other parameters based on the user's instructions. 
The resulting specification defines a customized optimization problem that affords a computational optimization for generating individualized, optimal UI solutions. 
Therefore, our agent overcomes a significant limitation of prior adaptive interface efforts, operating within the fixed optimization framework described in Sec.~\ref{sec:introduction} \& \ref{sec:personalization-auis}.

We define the general multi-objective optimization problem as:
\begin{equation}
\begin{aligned}
& \underset{\mathbf{x} \in X}{\min } \quad \{ f_1(\mathbf{x}, \mathbf{P}_1), f_2(\mathbf{x}, \mathbf{P}_2), \ldots, f_k(\mathbf{x}, \mathbf{P}_k) \} \\
& \text{subject to} \quad g_j(\mathbf{x}) \leq 0, \quad j = 1, 2, \ldots, m,
\end{aligned}
\label{eq:moo}
\end{equation}
where $\mathbf{x}$ is the decision variable within the design space $X$, $f_i$ is an objective function, and $g_j$ denotes a constraint.
The vector $\mathbf{P}_i$ represents the parameters of the respective objective function.
Therefore, our agent dynamically specifies the optimization problem during Configuration by including relevant objective functions $\{f_1,\ldots,f_k\}$ and setting the values of the corresponding parameters $\mathbf{P}_i$ based on the user's instructions.

\subsection{Optimization: Searching and generating Pareto-optimal layout solutions}

At this stage of the framework, a computational solver is executed to solve the optimization problem specified in the Configuration phase.
Our framework thus includes the parameterized subset of objective functions during Optimization and identifies the solutions to the specified MOO problem.
The resulting Pareto-front consists of solutions that reflect various combinations of prioritized objectives.
As a result, they represent a spectrum of optimal design candidates that are aligned with the user's expressed preferences.

%
Because the optimization may discover a large number of similar Pareto-optimal designs, a smaller set of representative designs to manage solution quantity and diversity is retained.
Each of the designs we keep reflects a distinct trade-off within the objective space.
We identify these designs using a scalarization-based filtering of the Pareto-optimal set (similar to~\citet{johns2023towards}).
We denote the final set of candidate solutions as $X^*$.


\subsection{Validation: Selecting the one layout solution that best fits the user's instructions}
\label{sec:validation}

In a final step, a third agent validates that the output of the optimizer aligns with the user's expressed preferences.
Identifying the final UI layout $\mathbf{x^*}$ from $X^*$ 
is challenging, because it requires assessing all solutions in $X^*$ in the context of the user's (expressed) preferences. 

We leverage the VLM-based agent's reasoning capabilities to evaluate the generated layouts in terms of the subjective instructions.
Our agent compares all layout candidates in $X^*$ against the user's initial instructions by prompting a custom VLM instance.
This prompt comprises the user's instructions, an image of their physical surroundings, and the optimized virtual widgets.
Our agent returns the layout $\mathbf{x^*} \in X^*$ that best aligns with the instructions. 

In summary, our agent-assisted selection from all Pareto-optimal results overcomes the previously needed manual effort~\cite{johns2023towards} during UI adaption.
Our framework achieves this by validating layout candidates and picking the one that is most suitable UI following a user's prompt and in the context of their environment.

\subsection{Aggregating user preferences over time}

In addition to obtaining the final UI layout $\mathbf{x^*}$, our framework also integrates user's expressed preferences over time.
As a user provides new instructions, our framework combines this new input with all previous instructions.
Aided by our Ambiguity Detection, this results in a comprehensive collection of instructions over time, which our agent leverages in every subsequent step.
This allows our agent to interpret the user's instructions and desired adjustments with increasing accuracy, while ensuring that the optimized layouts resulting from our framework align increasingly better with the user's overall preferences.

\begin{figure*}[t]
    \centering
    \begin{subfigure}[t]{0.3\textwidth}
        \centering
        \includegraphics[width=\textwidth]{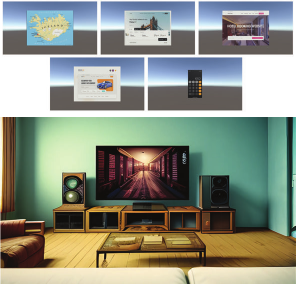}
        \caption{Task: plan a travel and book the hotel}
        \label{fig:data_collection_scenarios_subfig1}
    \end{subfigure}
    \hfill
    \begin{subfigure}[t]{0.3\textwidth}
        \centering
        \includegraphics[width=\textwidth]{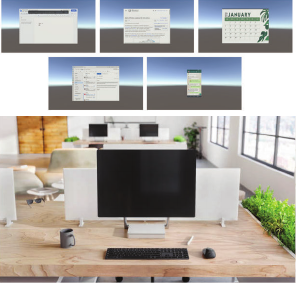}
        \caption{Task: write an article with information based on the Wikipedia}
        \label{fig:data_collection_scenarios_subfig2}
    \end{subfigure}
    \hfill
    \begin{subfigure}[t]{0.3\textwidth}
        \centering
        \includegraphics[width=\textwidth]{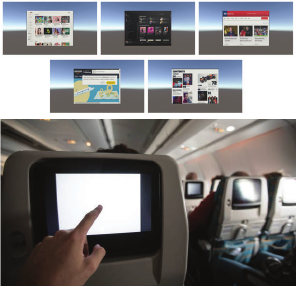}
        \caption{Task: watch sports videos}
        \label{fig:data_collection_scenarios_subfig3}
    \end{subfigure}%
    \vspace{-3mm}%
    \caption{Examples of scenarios from the data collection survey, showcasing virtual widgets and the simulated environment. 
    Participants are asked to imagine themselves in the given background and to provide instructions to an intelligent assistant to arrange the virtual widgets that completes the specified task. }
    \Description
    {
    The image displays three tasks, each with two accompanying images: Task (a): “Plan a travel and book the hotel.” The images show a laptop with a beach scene on the screen and a hand holding a credit card over another laptop displaying a booking website. Task (b): “Write an article based on Wikipedia.” The images include an open laptop with Wikipedia’s webpage visible and another screen showing an article being composed. Task (c): “Watch sports videos.” The images depict multiple screens showing various sports and someone pointing at a tablet or screen displaying sports content.
    }
    \label{fig:data_collection_scenarios}
\end{figure*}

\section{Example use case: Mixed Reality UI layout adaptation}
\label{sec:example-use-case}
To evaluate the effectiveness of our \textit{Auto-Optimization} framework in a representative use case, we apply it to the adaptation of user interface layouts in Mixed Reality. 
Unlike traditional interfaces confined to a single device, MR interfaces extend into the user’s physical environment, enabling content to be positioned around the user in spatially meaningful ways. 
However, the vast design space of potential layouts, combined with constraints imposed by the physical environment and various contextual factors, makes automatic UI adaptation essential (as evident by numerous research efforts, e.g., \cite{semanticadapt2021cheng,cheng2023interactionadapt,ens2015spatial}).

In this paper, we use this use case to concretely implement the previously conceptualized \projname{} framework, enabling us to evaluate its effectiveness and compare it to the current state of the art in UI optimization. 
In the following sections, we first describe the data collection process for fine-tuning the VLM-based agents, then present the use-case-specific implementation of our framework, and finally compare its performance to state-of-the-art approaches within the context of this use case.


\section{Data Collection}
\label{sec:data_collection_study}
To guide the adaptation of MR layouts based on verbal instructions, our approach requires relevant data.
Since no such dataset existed, we conducted an online survey to gather input from MR users on how they would provide verbal instructions to an intelligent assistant that could layout their MR interface.

\subsection{Survey Design}
The survey was designed in alignment with the core concept of our framework: intelligent agents interacting with an MR UI multi-objective optimization scheme to enable easy customization of the layout adaptation process. 
For the optimization, we defined several objectives identified from related work that our framework should include: physical body strain and exertion~\cite{johns2023towards}, haptic feedback for touch interactions~\cite{cheng2023interactionadapt}, UI design principles (such as alignment)~\cite{semanticadapt2021cheng}, and the interface's relation to the physical world (including semantic relationships and overlay suitability)~\cite{semanticadapt2021cheng}.

Before the survey, we explained to participants that their task was to provide instructions to an intelligent assistant capable of generating MR UIs based on those instructions. 
We explicitly outlined the objectives the assistant could consider and illustrated the process with two example instructions showing how the assistant could produce corresponding MR layouts. Participants then completed demographic questions before proceeding to the main survey.

\subsubsection*{Scenarios}

Our survey featured 15 distinct scenarios in which participants were asked to provide instructions. Each scenario included a first-person perspective photo depicting the surroundings of a hypothetical MR user, along with a specific task and a set of virtual widgets. 
We designed three distinct sets of virtual widgets, each tailored to a specific context: work, entertainment, and travel planning.
In each scenario, the task was designed to align with the selected widgets. Examples include scheduling a meeting (work widgets), online shopping (entertainment widgets), and booking a flight (travel planning widgets). 
The background photo was chosen to match the task setting, such as an office space, living room, or coffee shop.
\autoref{fig:data_collection_scenarios} showcases three examples from our survey, illustrating the tasks and virtual widgets for each scenario.

\subsubsection*{Questions.}
In each scenario, participants were asked to provide instructions to the intelligent assistant on how to adapt the widgets to the background such that they could best perform the specified task ("Please provide the instructions to the Intelligent Assistant, so it can adapt the UI based on the current context and task.").

\subsection{Participants}

We recruited 29 participants (19 female, 10 male), ages 19--43 (M=26, SD=3.42) from an online crowd-sourcing platform.
To guarantee a certain level of MR experience among participants, we screened them to ensure they used a MR device at least 1 to 5 times a month.
Of those, participants reported their frequency of using MR: three participants reported daily use, six mentioned using it several times a week, eleven indicated they used it several times a month, and the remaining participants used it less frequently.
Participants completed the survey in 30\,min and received \pounds 4 as a gratuity.

We excluded participants who answered one or more of our three control questions incorrectly.
Consequently, the data from 27 participants were used in the analysis.





\subsection{Formative Analysis}
From the 27 participants, we collected a total of 415 instructions, with an average length of 26.51 words (SD=31.56).
The goal of our analysis was to determine whether these instructions could be directly utilized by our intended framework to guide a multi-objective UI optimization with the specified objectives.
Three researchers independently coded all collected instructions.
They categorized the instructions into two groups: 
\begin{enumerate} 
\item Well-formed instructions that included all necessary information, or had missing details that could be inferred from the context. 
\item Incomplete or ambiguous instructions that lacked critical information which could not be inferred. \end{enumerate}
A majority vote among the three researchers was used to determine the final categorization of each instruction. 
As a result, 217 out of 415 responses were classified as well-formed, while the remainder were deemed ambiguous or incomplete.
%
%
These results are consistent with findings from previous research~\cite{mehrabi2023resolving, weerakoon2020gesture, hatori2018interactively} where participants also tended to provide incomplete instructions.
This supports the validity of our design choice to incorporate an ambiguity detection module aimed at identifying ambiguous or incomplete input.
\section{Use Case Implementation}
\label{sec:implementation}
In this section, we describe the implementation of the \projname framework for the specific use case of MR layout adaptation. 
While adhering to the general structure outlined in Section \ref{sec:framework}, we incorporate use case specific objective terms and constraints. 
Additionally, we leverage the data collected in Section \ref{sec:data_collection_study} to adapt the VLM instances for this application.

\subsection{Ambiguity Detection}
\label{sec:pipeline_ambiguity}


We adapt the VLM used in this phase through few-shot learning by augmenting its context with data from our collection study.
Specifically, we provide the module with all 405 pairs of collected instructions, along with their assigned categories (well-informed or ambiguous), to improve its understanding of the task.
\autoref{app:ambiguity-prompt} includes the comprehensive prompt that accompanies this module.

After the Ambiguity Detection module resolves all ambiguous aspects of the user’s instruction through clarification questions and the user’s additional responses, it concatenates these user input into a single, complete instruction and passes it to the Configuration module in natural language. 
Keeping the instruction in natural language is an intentional design choice: it preserves readability and explainability, making the subsequent configuration decisions transparent and easier for human users to interpret.



\subsection{Configuration}
As described above, this agent specifies the optimization problem for which solutions will be identified in the next step of our framework. 
To tailor it to our specific use case, we incorporate well-established objective functions, their respective parameters, and constraints drawn from the adaptive MR layout literature. 
The customized agent is then employed to determine the relevance of these components based on the user's verbal instructions. 
The following subsections provide a step-by-step explanation of this process.

\subsection*{\sscnt{a} Objective functions}

For the parameterizable optimization problem, we define the complete set of objective functions ${f_1, \ldots, f_k}$ using terms from prior work that have proven effective in generating usable MR layouts:

\begin{description}
\item[Spatial alignment~\cite{semanticadapt2021cheng}] 
For each virtual widget, identify the voxels it occupies and compare them with the voxels occupied by other widgets.
Calculate the distance between corresponding rows and columns of these voxels. 
A higher alignment cost indicates that the widget is poorly aligned with others, both vertically and horizontally.
A lower cost signifies that the widget is well-aligned with neighboring widgets across both dimensions.
    
\item[Field of view~\cite{semanticadapt2021cheng}]
For each virtual widget, calculate the angular difference between the widget's position and the user's gaze direction and multiply it with its observation probability. 
If the angular difference is less than 5 degrees (i.e., within the foveal region~\cite{millodot2014dictionary}), the cost is zero. 
A greater angular difference results in a higher cost, indicating that the widget is far from the user's gaze direction, while a lower cost reflects near to the foveal area.
    
\item[Anchor to physical object~\cite{evangelista2022auit}]
If a virtual widget is anchored to a physical object, anchoring cost stems from the angular distance between the virtual widget and the physical object from the user's perspective. 
A higher cost indicates that the virtual widget is more distant from its anchored object, whereas a lower cost suggests closer alignment. 
If the virtual widget is not anchored to any physical object, the anchoring cost is considered zero.
    
\item[Overlay physical object~\cite{li2024situationadapt}]
For each physical object, identify the voxels it occupies and check if any virtual widgets are positioned between these voxels and the user. 
If there is any, the overlaying cost is calculated as the angular differences between the virtual widget and the physical object, weighted by the object' overlay suitability. 
A higher overlaying cost indicates that physical objects that are not well-suited for overlay are obscured by virtual widgets, while a lower cost suggests a clearer view of the physical objects.
    
\item[Neck strain~\cite{johns2023towards}] 
For each virtual widget, calculate the angular distance between the user's eye and the widget's position on the x-z plane, and then multiply this distance by the widget's observation probability. 
A higher cost indicates that widgets with higher observation probability are positioned above or below the user's eye level, whereas a lower cost reflects widgets that are closer to the eye level.

\item[Arm exertion~\cite{johns2023towards}] 
Similar to the neck strain objective, for each virtual widget, compute the angular distance between the user's shoulder and the widget's position, and then multiply this distance by the widget's interaction probability. 
A higher cost indicates that widgets with higher interaction probability are positioned significantly above or below the user's shoulder level, whereas a lower cost reflects widgets that are closer to the shoulder level, thus enhancing ease of interaction.    
\end{description}

\subsection*{\sscnt{b} Fixed constraints}

In addition to the objective functions, our optimization problem consists of fixed constraints $g_j(\mathbf{x})$ to ensure the optimized UI meets practical requirements.
These include preventing occlusion, maintaining an appropriate field of view, and a reachable distance of elements.
Specifically, these constraints include:

\begin{description}
\item[Occlusion~\cite{evangelista2022auit}] 
For each virtual widget, identify the voxels it occupies and check if any other virtual widgets are positioned between these voxels and the user. 
If there is any, the occlusion constraint cost is positive; otherwise, the cost is zero. 
This ensures that all widgets remain visible to the user.
    
\item[Field of view~\cite{evangelista2022auit}] 
Similar to the field of view objective, each virtual widget must be within an angular distance of 60 degrees from the user to ensure it is within the user's field of view. 
If the angular distance exceeds this threshold, the distance constraint cost is positive; otherwise, the cost is zero.
    
\item[Distance~\cite{evangelista2022auit}]
We assume that all virtual widgets should be within arm's reach. To enforce this, we compute the distance between each virtual widget and the user's shoulder, ensuring it does not exceed 0.65. 
If the distance exceeds this threshold, a positive distance constraint cost is applied; otherwise, the cost is zero.
This threshold can be adjusted if direct touch interaction is not necessary.
\end{description}

\subsection*{\sscnt{c} Parameterized objective functions}

Our optimization problem includes the parameters of individual objective functions $\{\mathbf{P}_1,\ldots,\mathbf{P}_k\}$ ($\mathbf{P_i}$ represents all parameters associated with objective $f_i$), relevant virtual widgets, and physical objects so as to better compute objectives and constraints.
Specifically, these parameters include:

\begin{description}
\item[Interaction probability of a virtual widget~\cite{cheng2023interactionadapt}] 
For each virtual widget, estimate the probability that the user will interact with it using direct touch.

\item[Observation probability of a virtual widget~\cite{cheng2023interactionadapt}] 
For each virtual widget, estimate the probability that the user will observe it.

\item[Overlay suitability of a physical object~\cite{li2024situationadapt}]
For each physical object, estimate the suitability of overlaying a virtual widget over it. 

\item[Anchored to physical object for each virtual widget~\cite{evangelista2022auit}]
For each virtual widget, determine whether it is anchored to a physical object. If it is, identify the specific physical object to which it is anchored.
\end{description}

\begin{figure*}[t]
  \centering
  \includegraphics[width=\textwidth]{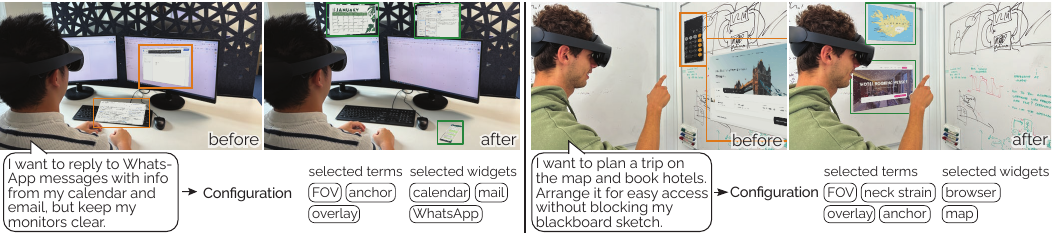}%
  \vspace{-1mm}%
  \caption{Illustrative demonstration of our Configuration component in a Mixed Reality context. 
  Left: The user aims to reply to messages using information from virtual widgets while avoiding overlaying physical monitors. 
  The Configuration module in our framework selects the objective terms that relate to the user's expressed preferences, such that the Optimizer produces layouts that keep all widgets within the user's field of view, the messenger anchored to the desk for haptic feedback, and the monitors visible and unoccluded.
  Right: The user plans a trip using a map and books a hotel while keeping the whiteboard sketch unobstructed. 
  Our framework selects and displays the widgets relevant to the user's instructions on the clean whiteboard.
  }
  \Description{The image displays a four-panel sequence showing a person at a workstation with two monitors, before and after implementing organization strategies. In the first ‘before’ panel, the individual is seen facing a monitor with an obscured screen due to multiple post-it notes attached to it. The second ‘after’ panel shows the same setup but with a clear monitor screen and post-it notes organized on the wall beside it. The third ‘before’ panel depicts another monitor blocked by books standing in front of it, while the fourth ‘after’ panel shows the books neatly arranged on shelves to the side, allowing an unobstructed view of the monitor. Text bubbles indicate that before organizing, tasks such as replying to WhatsApp messages and checking calendar events were hindered by cluttered monitors; after reorganizing, there is easy access without blocking important screens.}
  \label{fig:mrusage}
\end{figure*}

\subsection*{\sscnt{d} Configuring the overall objective function}

Based on the user's instructions,
our Configuration module specifies the optimization problem by selecting a subset of $\{f_1,\ldots,f_k\}$ and setting the values of the corresponding parameters $\mathbf{P}_i$.
To this end, we provide the respective agent with contextual information explaining the roles of objective functions, widgets, and their parameters. 
Based on this context, the agent first identifies the intended virtual widgets from a predefined set of candidates, thereby effectively defining a new design space $Y$ with $dim(Y) \leq dim(X)$.

For example, in the left scenario illustrated in~\autoref{fig:mrusage}, the Configuration module selects the "Calendar," "Mail," and "WhatsApp" widgets based on the user's instructions, determining the set of virtual elements to be optimized.
Similarly, based on the textual description of the available objective functions ($f_1,\ldots,f_k$) that we provide, the module selects the subset of objectives $\{ f_l(\mathbf{x}, \mathbf{P}_l), \ldots, f_n(\mathbf{x}, \mathbf{P}_n) \}$ where $1 \leq l \leq n \leq k$, that best represent the adaptation goals expressed by the user.
For instance, in the same scenario illustrated in~\autoref{fig:mrusage}, the module selects the "FOV," "Anchor," and "Overlay" objectives based on the user's instructions, ensuring that the optimization process prioritizes the most relevant adaptation criteria.
Finally, we provide the module with an explanation of the associated parameters ($\mathbf{P}_l,\ldots,\mathbf{P}_n$) of the selected subset of objective functions.
Since selecting appropriate parameter values depends on the characteristics of both the virtual widgets and the physical objects, we also provide the agent with the relevant properties.
The agent then estimates the values $\mathbf{P}_i$ for $i = l, \ldots, n$ to best reflect the user's input as captured by the previous module.
In the same scenario shown in the left of~\autoref{fig:mrusage}, the module determines a very low overlay suitability for the physical monitor, as the user prefers to keep their monitors clean.
Similarly, in the right scenario, it selects the left blackboard as the anchor target for the virtual widget, since the user wants easy access to the widgets while avoiding obstruction of their sketch.

The Configuration module outputs its optimization settings in JSON format, which the Optimization module then reads and uses to configure the optimizer accordingly.
\autoref{app:reasoning-prompt} prints the full prompt that instructs the agent of our Configuration module.


\subsection{Optimization and Validation}
Neither the Optimization module nor the Validation module needs to be specifically tailored to a particular application.
The Optimization module is responsible for solving the optimization problem that was adapted to the specific use case by the Configuration module.
It then solves the optimization problem, obtains a set of Pareto-front solutions, and sends them to the Validation module in JSON format.
The Validation module receives the solution set and parses each candidate. 
It then applies each configuration to the UI layout and captures corresponding screenshots of the interface and physical context, all while keeping these intermediate states invisible to the user to avoid disrupting the interaction experience.
It also receives the clarified user instruction to support subsequent evaluation.
The Validation module then selects the most suitable solution by assessing both its alignment with the user’s instruction and its overall usability.
\autoref{app:comparison-prompt} details the comprehensive prompt for the Validation module.





\subsection{Technical Implementation}
We implemented our system for the Meta Quest 3 using Unity 2021. 
The optimization module uses the AUIT framework~\cite{evangelista2022auit}, a toolkit designed to create adaptive Mixed-Reality applications.
The toolkit integrates directly with Unity, using Unity GameObjects and their properties as inputs for optimization.
We employ NSGA-III~\cite{blank2019investigating} to discover a set of optimal solutions on the Pareto-front based on the selected objectives (population size: 100, 40 generations).
The solver is implemented in Python 3.9 using the pymoo~\cite{blank2020pymoo} package and interacts with the Unity-based adaptation system. 
We apply the AASF decomposition\cite{wierzbicki1982mathematical} as the scalarization function to select a subset of solutions on the Pareto-front that balance different objectives effectively.
The process is initialized with a set of reference vectors, computed via Riesz s-Energy\cite{blank2020generating}, to ensure evenly distributed objective weight combinations.
We control the size of the reduced candidate set by adjusting the number of reference vectors.
We leveraged the GPT4 Vision 2024-05-01-Preview model of Azure OpenAI as VLM and access it via its Python API.
The individual modules communicated via socket-based networking.

\begin{figure*}[!h]
  \centering
    \includegraphics[width=\textwidth]{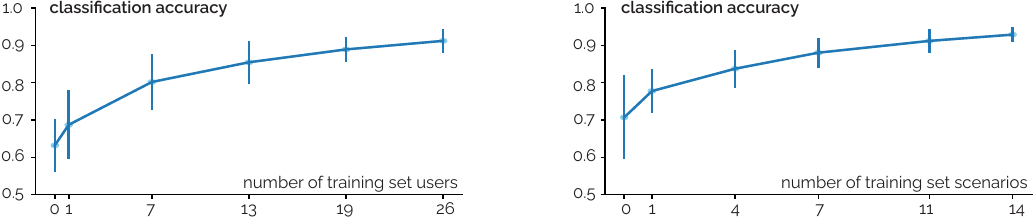}%
  \vspace{-2mm}%
  \caption{
  Classification accuracy of few-shot optimized VLMs with different number of participants and scenarios in training.}
  \label{fig:ambiguity_ablation}
  \Description{
  Two line graphs displaying the classification accuracy as a function of the number of training set users and the number of training set scenarios respectively. In the first graph, the horizontal axis is labeled “number of training set users” and ranges from 0 to 25. The vertical axis is labeled “classification accuracy” and ranges from 0.5 to 1.0. In the second graph, the horizontal axis is labeled “number of training set scenarios” and ranges from 0 to 14. The vertical axis is labeled “classification accuracy” and ranges from 0.5 to 1.0. Both graphs show a generally increasing trend, with error bars indicating variability at each data point.
  }
\end{figure*}

\section{Evaluation}
The objective of \projectname~is to streamline UI adaptation by automatically interpreting user instructions to define the optimization problem, specify inputs, and select suitable solutions.
To properly assess its effectiveness, we evaluate it in an MR layout adaptation use case.
To this end, we first assess individual modules of the pipeline. 
Specifically, we examine whether our system can successfully detect ambiguities in user prompts and whether the comparison module selects optimal MR layouts that align with those chosen by MR users. 
Finally, we assess the overall usability by comparing our end-to-end pipeline with baseline adaptation methods in a user study.

\subsection{Ambiguity detection}
To assess if \projectname~ can accurately detect ambiguities in user prompts, we utilized the data collected from our online survey to evaluate the performance of the ambiguity detection module.

Given that instructions could vary significantly across participants, we aimed to assess the scalability of few-shot learned VLMs, which serve as the underlying model of the Ambiguity Detection agent, when applied to new users. 
To achieve this, we performed a leave-{1, 7, 13, 19, 26, 27}-user-out cross-validation of the few-shot learned VLMs.
As outlined in Section \ref{sec:pipeline_ambiguity}, the instructions and their corresponding categories from the training set were incorporated into the prompt. 
A VLM instance was then asked to categorize the instructions in the test set, and the classification results were compared to the researchers' decisions. 
Figure \ref{fig:ambiguity_ablation} (left) illustrates the results.
Leaving all users (27) out resulted in an accuracy of 63.22\%~(SD = 7.20\%), while including data from 7 users in the training set and leaving out 20 users resulted in an accuracy of 80.23\%~(SD = 7.40\%), which significantly improved classification accuracy.
With leave-one-user-out, the few-shot learned VLMs achieved a classification accuracy of 91.26\%~(SD = 4.69\%).
These results show that while the VLMs improve with additional user data, they already deliver strong performance using data from just half of the users.

Because users’ instructions vary substantially across scenarios, we also investigate the capability of few-shot learned VLMs to detect ambiguity in previously unseen situations.
With the instructions collected from 15 scenarios, we performed a leave-{1, 4, 8, 11, 14, 15}-scenarios-out cross-validation. 
We incorporated user prompts and their categories from the training scenarios and asked the VLMs to categorize instructions in the test scenarios, which simulated the detection of ambiguity in practical settings.
The results are illustrated in the right panel of Figure \ref{fig:ambiguity_ablation}.
When all scenarios were excluded from the training set, the VLMs relied solely on their intrinsic capabilities for ambiguity detection, achieving an accuracy of 70.73\% (SD = 11.28\%). 
Incorporating data from 7 scenarios into the training set significantly improved accuracy to 88.09\% (SD = 4.06\%), and when only 1 scenario was excluded, the VLMs achieved an accuracy of 92.93\% (SD = 2.05\%).
These results indicate that few-shot learned VLMs can effectively detect ambiguity in unseen scenarios by learning from a limited number of training scenarios.

\subsection{Layout comparison}
\label{sec:comparison_validation}

A key feature of our pipeline is its ability to enable textual guidance for the Pareto front exploration when searching for the optimal layout candidate.
The assumption behind this feature is that \projectname will evaluate the alignment between textual layout descriptions and layout candidates in a manner similar to human users.
To assess this, we conducted an online survey to compare how experienced MR users and \projectname rate the alignment of verbal descirptions and generated MR layouts.

\subsubsection{Survey design}
The survey was designed to evaluate how \projectname and experienced MR users rate the alignment of textual layout descriptions with virtual UIs in different scenarios. 
Prior to the survey, participants were briefed on how \projectname functions and how the MR layouts were generated. 
After this introduction, participants provided demographic information before proceeding to the main survey.

\paragraph{Scenarios}
The main part of our survey included 18 scenarios. 
Each scenario consisted of four images of MR layouts rendered in the same environment, which were generated by \projectname based on text prompts provided by participants of our data collection study.
The text prompts were drawn from three specific environments: an office space, a living room, and an airplane.

To ensure consistency, we recreated the environments from our data collection study in Virtual Reality and used these replicas to render the MR layouts. 
Participants were instructed to treat the VR environments as real-world physical surrounding, with only the MR layouts were to be considered virtual.

\begin{figure*}[t]  
    \centering
    
    \begin{subfigure}[h]{\textwidth} 
        \centering
        \includegraphics[width=\textwidth]{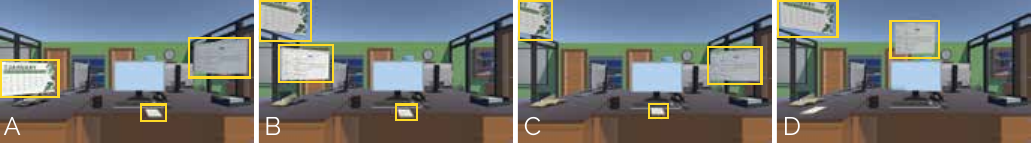}
        \caption{Scenario 8 in an office generated using the prompt: "I'd like to schedule a meeting, open the calendar widget on the left and the email widget on the right of the laptop, and the messenger widget under the laptop, don't occlude the laptop." Design (a) is preferred by \llms{} and \ptps{}.}
    \end{subfigure}
    
    \begin{subfigure}[h]{\textwidth}
        \centering
        \includegraphics[width=\textwidth]{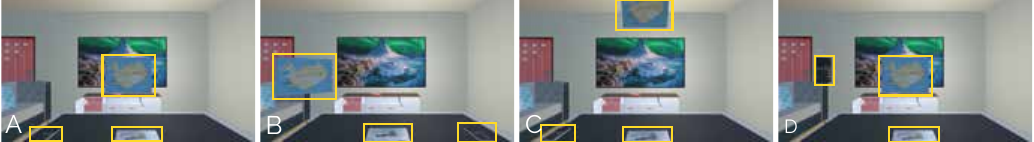} 
        \caption{Scenario 4 in a living room generated using the prompt: "Place the map on the tv and put the flight booking and calculator on the desk so I can interact with it." Design (a) is preferred \llms{} and \ptps{}.}
    \end{subfigure}
    
    \begin{subfigure}[h]{\textwidth}
        \centering
        \includegraphics[width=\textwidth]{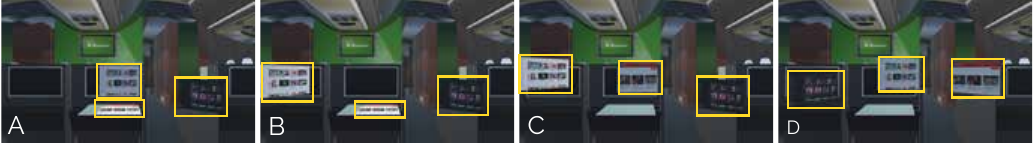} 
        \caption{Scenario 16 in an airplane generated using the prompt: "I would like to watch videos on the tv and scroll and interact with the sports news, maybe play music at the same time, please place the spotify next to youtube." Design (d) is preferred by \llms{} and \ptps{}.}
    \end{subfigure}
    \vspace{-10pt}

    \caption{Examples of scenarios from the layout comparison survey, each displaying four candidate solutions (a–d), the prompt it was generated with, and the design preferred by participants of the survey. The scenario numbers and candidate solutions (a–d) correspond to the respective scenarios and categorical values (A–D) displayed in Figure \ref{fig:comparison_results}.
    }
    \Description{The image presents three scenarios, each with a sequence of four pictures labeled A, B, C, and D. Each sequence illustrates different user interface designs within a virtual environment. Scenario 1: Depicts an office setting with various digital widgets on a desk and laptop screen. The task is to schedule a meeting using the calendar widget on the left and the email widget on the right of the laptop, avoiding the messenger widget under the laptop. Design (a) is preferred by VLMS and PPS. Scenario 2: Shows a living room environment where virtual objects are placed on a physical desk. The task is to place the map on the TV pad and put the flight booking and calculator on the desk for interaction. Design (a) is preferred by VLMS and PPS. Scenario 3: Displays another office setting with different arrangements of digital widgets around a physical desktop monitor. The task prompt is: "I would like to watch videos on the tv and scroll and interact with the sports news, maybe play music at the same time, please place the spotify next to youtube.}
    \label{fig:comparison_scenarios}
\end{figure*}

The layouts represented Pareto-optimal design candidates generated by Pareto-AUIT based on input from our Configuration module. 
These selected solutions were rendered in VR, and screenshots were captured.
Figure \ref{fig:comparison_scenarios} illustrates the four candidate designs along with the corresponding text prompt for three sample scenarios, each rendered in one of the target environments.

\paragraph{Question.}
For each scenario, participants should select the layout that best aligned with the prompt from the four options provided (``Please select the layout that best aligns with the user's instructions.'').
The question was presented as alternative forced choice with four stimuli (4-AFC)~\cite{yamamoto2022photographic}.


\begin{figure*}[t]
  \includegraphics[width=\textwidth]{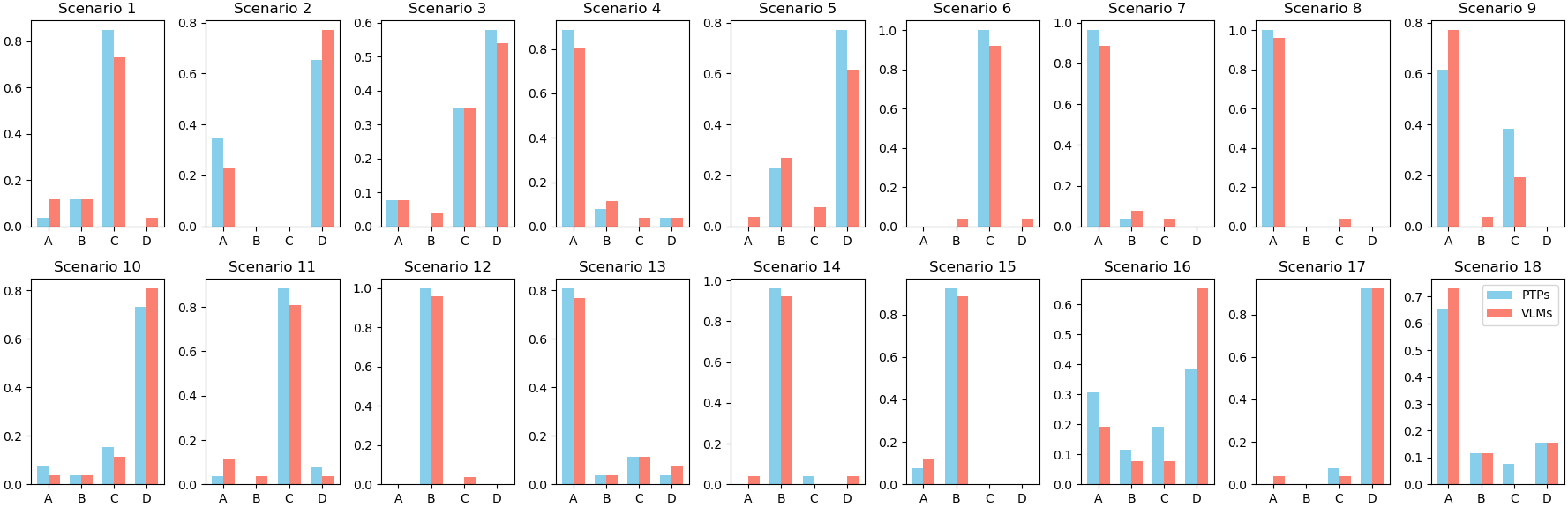}%
  \vspace{-2mm}%
  \caption{Categorical distributions of the MR layout voted as best aligned by \llms{} and \ptps{} in each scenario.}
  \Description{The image displays a series of 18 bar graphs, each labeled as ‘Scenario’ followed by numbers 1 through 18. Each graph contains bars in two different colors, red and blue, representing different categories labeled A through F. The vertical axis of each graph ranges from 0 to 0.6, indicating the frequency or proportion of each category within that scenario. The graphs are arranged in a grid pattern with two rows and nine columns.}
  \label{fig:comparison_results}
\end{figure*}

\subsubsection{Participants}

We recruited 27 participants (15 female, 12 male), ages 18--54 (M=35, SD=10.8) from an online crowd-sourcing platform.
To guarantee a certain level of VR experience among participants, we screened them to ensure they used a VR device at least 1--5 times a month.
Of those, one participant reported using VR 11--15 times a month, 3 participants used it 6--10 times, and the remaining participants used it 1--5 times per month.
Participants completed the survey in around 20\,min and received \pounds 3 as a compensation.

We excluded one participant who answered two out of three control questions incorrectly.
Consequently, the data from 26 participants (\ptps{}) were used in the analysis.

\subsubsection{VLM ratings}

We used the same scenarios, text prompts, and layout images to generate ratings with \projectname. 
Each scenario's four images and corresponding textual prompt were fed into \projectname, which returned its best layout (using the approach described in \autoref{sec:validation}).
To ensure a balanced comparison, we generated results from 26 distinct VLM instances of the agent of the Validation module, matching the sample size of the human participants.
For our analysis, this yielded a total of 468 ratings per condition (\llms{} and \ptps{}).

\subsubsection{Results}

The goal of our analysis was to determine if \projectname assesses alignment between verbal instruction and generated MR layout similar to the population of MR users.
To this end, we computed the categorical distributions of the MR layout voted as best aligned by \llms{} and \ptps{} in each scenario. 
As the categorical data is discrete, we analyzed the relationship between these distributions using the Chi-Squared Test.
Our results suggested that there was no difference between the distributions of the ratings by \llms{} and \ptps{} in any of the scenarios ($p > 0.05$).

In terms of the validity of our approach, an essential factor is whether the mode of the categorical distributions from both groups overlaps. 
We calculated the fraction of scenarios where the mode was consistent between \llms{} and \ptps{}, and found that this was the case for all scenarios, yielding a perfect fraction of $1.0$. 
A high degree of overlap and generally low divergence between the distributions can further be observed in \autoref{fig:comparison_results}.
\autoref{fig:comparison_scenarios} shows three examples of the four candidate designs, highlighting the one preferred by both \llms{} and \ptps{}. The scenario numbers and candidate solutions (A–D) correspond to the respective scenarios and categorical values (A–D) displayed in \autoref{fig:comparison_results}.


\subsection{User study}
\label{sec:evaluation}

To evaluate the usability of \projname, we conducted a user study comparing it to two baseline methods: manual widget placement and choosing a layout from the Pareto front.

\subsubsection{Design}
We employed a within-subject study design with one independent variable, \ivmethod{}, which had three levels: \cours{}, \cpareto{}, and \cmanual{}.
To control for potential influence of environments, we introduced a control variable, \ivscenario{} (with three levels: living room, office, and airplane), representing the context in which participants were observed. 
In each condition, participants were asked to customize their user interface layout using each \ivmethod{} in all three \ivscenario{}s.
They specified what they would like to do with the given widgets to the system first and then adjusted the generated user interface layouts.
For dependent variables, we recorded the \emph{number of adjustments} made to the user interface and the \emph{adjustment distance} between the initial system-generated layout and the final layout refined by participants. 
We also employed the NASA-TLX questionnaire~\cite{hart1988development} to assess users' workload using each \ivmethod{}.
Additionally, participants were asked to rank screenshots of the \cmanual{}-adjusted interfaces, as well as the \cours{}- and \cpareto{}-generated user interfaces (prior to any manual adjustment) at the end of each scenario.
The order of both \ivscenario{} and \ivmethod{} was counterbalanced using a Latin square.

\subsubsection*{Scenarios}

The same three virtual environments as in the layout comparison study were used: a living room, an office, and an airplane (as illustrated in \autoref{fig:comparison_scenarios}).

\subsubsection*{Methods}
We compared \cours{} to both \cbaseline{}~\cite{johns2023towards} and \cmanual{}.
\begin{description}
\item[\cours{}] 
Participants were asked to specify the task they aimed to complete with the virtual widgets in the current scenario and then provide instructions for the desired user interface layout.
If the ambiguity detection module raised any clarification questions, participants responded accordingly. 
After our approach generated the MR layout, participants reviewed the interface and, if desired, refined it via manual drag-and-drop of widgets.
    
\item[\cpareto{}] 
With \cbaseline{}, we modeled the optimization problem with pre-defined widgets, objective goals, and parameters.
After the optimization finished, participants were presented with up to four candidate layouts from the Pareto front, selected using the AASF decomposition scalarization function, and asked to choose the one that best aligned with their preferences.
They had the option to refine the layout via widget drag-and-drop.

\item[\cmanual{}] 
To begin, all virtual widgets were arranged in front of the user. 
Three widgets were placed at eye level, while the remaining two were positioned at hand level. 
This layout is comparable to the typical virtual environments on commercial platforms like Meta Quest and Apple Vision Pro.
The order of the virtual widgets was randomized. 
Participants were given the option to modify the initial layout as needed.
They could adjust the position and depth of the virtual widgets as needed and hide widgets by pressing a button on the controller.
\end{description}

\subsubsection{Procedure}
Participants started the study by providing informed consent and completing a demographic questionnaire. 
Following this, they engaged in a training trial designed to familiarize them with the available user interface elements and the experimental environment. 
During the training phase, participants were introduced to the workflow of all three \ivmethod{}s.
Subsequently, participants completed the conditions of the study, during which they provided instructions and observed the generated layouts across three scenarios for each method, resulting in a total of nine trials. 
After each trial, participants were asked to rank the layouts generated by each method.
Naturally, the ranking of \cmanual{} took place after the final layouts were created manually by the participants. 
For \cours{} and \cbaseline{}, the participants provided ranks before each final manual layout tuning; We chose to remove the factor of manual tweaking because our primary goal is to compare the quality of the layouts produced by these two methods against fully manual placement.
After each condition, participants were asked to complete the NASA-TLX questionnaire.
The entire study session was completed within 40 minutes.

\subsubsection{Participants}
We recruited 12 participants (4 female, 8 male), ages 22--28 (M=25.91, SD=1.68) from a local university.
They reported their frequency of VR/AR headset use: four indicated using it several times a week; eight reported using it several times a month.

\begin{figure*}[t]
  \centering
  \includegraphics[width=\textwidth]{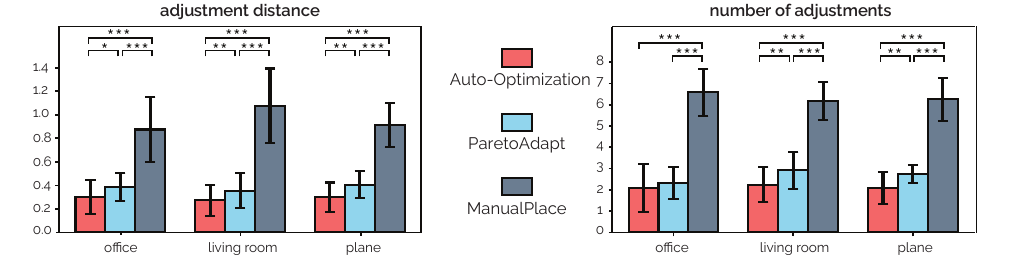}%
  \vspace{-3mm}%
  \caption{
  Means and standard deviations of the distance and the number of adjustments for each condition across all scenarios\\(*: $p < 0.05$, **: $p < 0.01$, ***: $p < 0.001$).}
  \Description{
        The image contains two bar graphs side by side. The left graph is labeled ‘adjustment distance’ and the right graph is labeled ‘number of adjustments.’ Both graphs have three pairs of bars, each pair corresponding to a different setting: office, living room, and plane. In both graphs, one bar in each pair is red, the other blue and the third one grey, with a legend indicating red represents ‘AutoOptimization’, blue represents ‘Pareto-Adapt.’, and grey represents ‘Manual Placement’. Each bar has an error bar indicating variability.
  }
  \label{fig:eva_results}
\end{figure*}

\subsubsection{Apparatus}
Participants wore a Meta Quest 3 headset, tethered to a desktop computer equipped with an Intel Core i7-12700K processor, an NVIDIA GeForce GTX 3070 GPU, and 32 GB of RAM. 
They interacted with the application using both a keyboard and the Meta Quest 3 controllers.

\subsubsection{Results}

We analyzed the effect of \ivmethod{} on three variables: \emph{adjustment distance}, \emph{number of adjustments}, and NASA-TLX scores, across all \ivscenario{}s.
Since the assumptions of normality (Shapiro-Wilk \pvall{<}{.05}) and sphericity (Levene's \pvall{<}{.05}) were violated for both \emph{adjustment distance} and \emph{number of adjustments}, we conducted a one-factor Aligned Rank Transform (ART) ANOVA for these two variables as well as to the NASA-TLX measures.

\paragraph{User interface adjustments}
We found a main effect of \ivmethod{} on both \emph{adjustment distance} \anova{2}{99}{98.75}{<}{.001}~ and \emph{number of adjustments} \anova{2}{99}{112.05}{<}{.001}~ across \ivscenario{}s (Figure \ref{fig:eva_results}). 
Post-hoc tests revealed that participants made fewer adjustments \ttest{11}{2.60}{<}{.05} and covered a shorter distance \ttest{11}{3.16}{<}{.01} before reaching satisfaction with layouts generated by \cours{} compared to \cbaseline{}. 
For \cmanual{}, participants began with a fixed central layout and manually adjusted all virtual elements, which could take multiple iterations. 
This process resulted in significantly higher adjustment counts and distances compared to both \cours{} \ttest{11}{-13.44}{<}{.001} and \cbaseline{} \ttest{11}{-10.28}{<}{.001}.
Appendix~\ref{app: layouts} shows selected participant-generated UI layouts.

\begin{figure}[b]
  \centering
  \includegraphics[width=\linewidth]{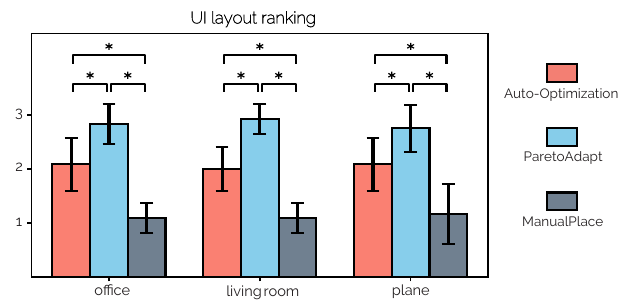}
  \vspace{-3mm}
  \caption{
  Ranking results of the layouts generated by the respective methods (lower is better). (*: $p < 0.05$, **: $p < 0.01$, ***: $p < 0.001$).}
  \Description{
  The image contains one bar graph with three pairs of bars, each pair corresponding to a different setting: office, living room, and plane. The graph is labeled ‘UI layout rankling.’ In each pair one bar is red, the other blue and the third one grey. Red represents ‘AutoOptimization’, blue represents ‘Pareto-Adapt.’, and grey represents ‘Manual Placement’. Each bar has an error bar indicating variability.
  }
  \label{fig:eva_ranking}
\end{figure}

\paragraph{Layout preference}
We observed a significant main effect of \ivmethod{} on participants' preference rankings \friedman{2}{53.56}{<}{.001} (Figure \ref{fig:eva_ranking}). 
Post-hoc test results indicated that \cmanual{} $(M=1.11, SD=0.39)$ outperformed both \cours{} $(M=2.05, SD=0.47)$ \wilcoxon{64.5}{<}{.001}~ and \cpareto{} $(M=2.83, SD=0.37)$ \wilcoxon{4.5}{<}{.001}~, while \cours{} was ranked higher than \cpareto{} \wilcoxon{87.5}{<}{.001}~.
Participants always ranked the layouts generated through \cmanual{} as their top choice. 
This is expected, as manual adjustment allowed them to directly position each widget to perfectly align with their implicit preferences. 
It is worth noting that layouts generated by \cours{} were consistently ranked higher than those generated by \cpareto{}, highlighting that our approach generates layouts that are better aligned with the users' preferences.
Participants attributed their preference to the fact that, although \cbaseline{} provided several reasonable options, these did not closely align with their ideal layouts. 
In contrast, \cours{} enabled participants to provide specific instructions, resulting in layouts that better matched their expectations.

\begin{figure*}[t]
  \centering
  \includegraphics[width=\textwidth]{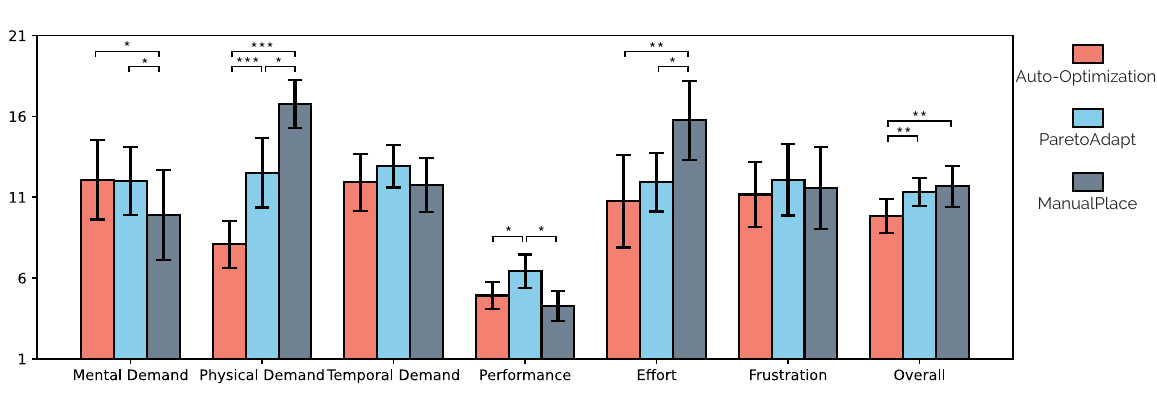}
  \vspace{-5mm}
  \caption{Means and standard deviations of the NASA-TLX questionnaire results for all three methods. 
  (*: $p < 0.05$, **: $p < 0.01$, ***: $p < 0.001$).
  }
  \Description{
  The figure is a bar graph comparing three groups across seven workload dimensions: Mental Demand, Physical Demand, Temporal Demand, Performance, Effort, Frustration, and Overall workload. Each group is represented by a different colored bar—red, light blue, and dark gray—with error bars indicating variability. Red represents ‘AutoOptimization’, blue represents ‘Pareto-Adapt.’, and grey represents ‘Manual Placement’. In the Mental Demand category, both the red and light blue groups reported significantly higher scores than the dark gray group. For Physical Demand, there are significant differences between all three groups: the dark gray group had the highest ratings, followed by the light blue group, with the red group reporting the lowest demand. In Performance, the light blue group rated themselves significantly higher than the red and dark gray groups, indicating better perceived performance. When it comes to Effort, the dark gray group reported the highest effort, significantly more than both the red and light blue groups. For Overall workload, both the light blue and dark gray groups had significantly higher scores compared to the red group. No significant differences were found among the groups for Temporal Demand or Frustration.
  }
  \label{fig:eva_sub_results}
\end{figure*}

\paragraph{Perceived workload}
Figure \ref{fig:eva_sub_results} illustrates the results of the NASA-TLX.
Our analysis revealed a main effect of \ivmethod{} on \emph{Mental Demand} \anova{2}{33}{3.46}{<}{.05}{}, \emph{Physical Demand} \anova{2}{33}{75.89}{<}{.001}{}, \emph{Performance} \anova{2}{33}{14.94}{<}{.001}{}, \emph{Effort} \anova{2}{33}{11.35}{<}{.001}{}, and \emph{Overall} workload \anova{2}{33}{8.47}{<}{.005}{}.
Post-hoc tests indicated that while participants perceived higher mental demand with \cours{} compared to \cmanual{} \ttest{11}{2.88}{<}{.05}, they also reported lower physical demand \ttest{11}{-12.30}{<}{.001} and effort \ttest{11}{-4.46}{<}{.001}. 
Despite the increased mental demand, participants perceived no difference in performance using \cours{} and \cmanual{} \ttest{11}{1.82}{>}{.05}.
Furthermore, \cours{} resulted in the lowest overall workload score compared to \cpareto{} \ttest{11}{-3.23}{<}{.01} and \cmanual{} \ttest{11}{-3.82}{<}{.005}. 
These results suggested that while \cours{} required more cognitive engagement, it alleviated the overall workload of customizing user interfaces while being perceived similarly effective.

\paragraph{Task completion time}
We also analyzed the task completion time across \ivmethod{}s.
The results indicate that participants did not spend significantly more time using \cours{} $(M=46.83s, SD=12.03)$ than \cpareto{} $(M=44.52s, SD=5.38)$ \ttest{11}{-1.45}{>}{.05}, although completion times were notably longer compared to \cmanual{} $(M=11.47s, SD=2.07)$ \ttest{11}{13.05}{<}{.001}. 
The additional time required for \cours{} can be attributed to participants needing to manually input instructions via a keyboard. 
This process could be significantly expedited through the integration of more efficient interaction modalities, such as speech-based input.
In the case of \cpareto{}, the extended completion time is primarily due to its optimization process, which requires substantially more time than that of \cours{}. 
This is because \cpareto{} considers multiple objectives and explores a larger search space. 
However, despite the increased computational effort, it fails to produce results that better align with participants' customized needs, as evidenced by the greater number of UI adjustments required compared to \cours{}.
Furthermore, NASA-TLX responses indicated that participants did not perceive a higher time demand when using \cours{} compared to \cpareto{} and \cmanual{}, suggesting that the system's time consumption was acceptable from a user experience perspective.
Appendix~\ref{app: time_cost} reports the runtime of each module of our system.

\begin{figure*}[h]
    \centering
    \begin{subfigure}{\textwidth}
        \centering
        \includegraphics[width=\textwidth]{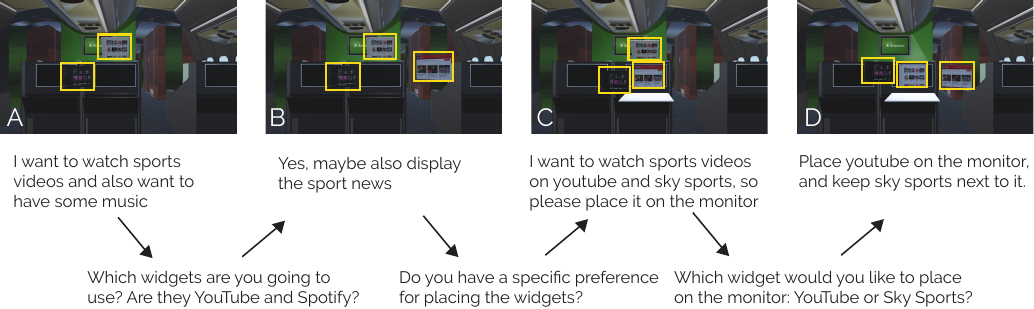}
        \caption{When the user provides a simple instruction about their task, our framework can only infer which widgets the user might want (A). 
        The framework then asks a clarification question to determine the exact widgets needed. 
        With the precise widgets identified, the framework assumes that the monitor in front of the user is unsuitable for overlaying widgets (B). 
        Consequently, it places the widgets around the monitor and asks the user to clarify their preferences for widget placement. 
        If the user specifies that the widgets should be placed on the monitor (C), it remains ambiguous which widget the user wants to overlay on the monitor. 
        Once this is clarified, the final layout (D) can be rendered.    \vspace{3mm}
        }
        \label{fig:ambiguity_comparison1}
    \end{subfigure}
    \hfill
    \begin{subfigure}{\textwidth}
        \centering
        \includegraphics[width=\textwidth]{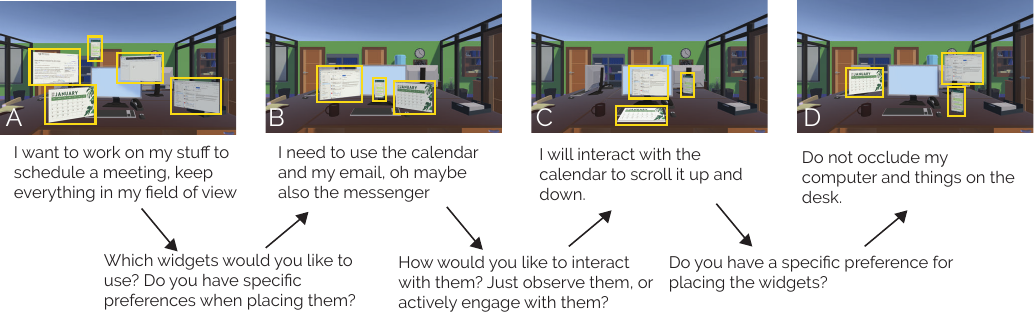}
        \caption{While the user does not specify the exact widgets to be used, our framework places all widgets by default (A) and asks for clarification on both the widgets and their placement preferences. After receiving responses, the framework generates layouts with the precise widgets (B), but further clarification is needed on how users would like to interact with them. Based on this response, the framework can infer the objectives related to haptic feedback and physical strain, and optimize the layout accordingly (C). As the user further clarifies their preferences, the framework can avoid overlaying certain physical objects and provide a customized layout (D).}
        \label{fig:ambiguity_comparison2}
    \end{subfigure}
    
    \caption{Examples of layouts generated from the user's instructions, which were partially ambiguous or incomplete.
    Following answers to our framework's clarification questions, the generated layouts were better aligned with the user's instructions (a--d).
    Note: Only the final layout (d) is presented to the user, whereas results (a--c) are only generated as examples.}
    \Description{The image is a flowchart with eight boxes labeled from A to H, each containing an illustration and text describing different scenarios involving the use of widgets on a computer monitor. The flow starts at box A with the question ‘I want to watch sports videos and also want to have some music services on my monitor. Which widgets are you going to use?’ It then branches off into two paths based on yes or no responses, leading through various decision-making processes about widget preferences, placement, and interactions. Each box provides a different context or question related to customizing the user’s experience with widgets on their monitor.}
    \label{fig:ambiguity_comparison}
\end{figure*}

\subsubsection{Ablation study: impact of disambiguating prompts}
\label{sec:ablation_ambiguity}
We conducted an ablation study to examine whether clarifying ambiguous prompts improved the quality of generated layouts. 
To this end, we generated intermediate results using participants' instructions before clarification, as illustrated in~\autoref{fig:ambiguity_comparison}.

In \autoref{fig:ambiguity_comparison}(a)A, the participant stated, "I want to watch sports videos and also want to have some music."
While there were two virtual widgets related to sports videos (Sky Sports and YouTube), the Configuration module selected only YouTube without ambiguity clarification. 
This decision conflicted with the participant's preference for Sky Sports, which better aligned with their intent to watch sports-specific content.
By answering the clarification question, the user indicated the widgets they preferred, resulting in the layout shown in \autoref{fig:ambiguity_comparison}(a)B.

Similarly, in \autoref{fig:ambiguity_comparison}(b)C, the user did not specify a clear objective, stating, "I will interact with the calendar to scroll it up and down."
Without ambiguity clarification, the system considered placing virtual widgets only on monitors and desks to provide haptic feedback for users. 
The clarification question prompted the user to provide more details about their objective. 
And when the user specified a preference for keeping the monitor and desk unoccluded, the optimizer placed the widgets in mid-air, which better accommodated the user's updated preferences.

Another type of ambiguity is illustrated in \autoref{fig:ambiguity_comparison}(a)C, where the participant said, "I want to watch sports videos on YouTube and Sky Sports, so please place it on the monitor."
The ambiguity arose from the user's use of "it," which left unclear whether they intended to place sports videos or YouTube on the monitor. 
This led the Configuration module to misinterpret the user's intent. 
After our ambiguity module prompted clarification, the user provided the necessary information, resulting in a layout that aligned with their preference.

\section{Discussion}
The goal of \projname is to enable users to customize their UIs using natural language by defining the optimization problem and selecting the optimal layout from a set of Pareto front solutions. 
In this section, we discuss the results of our evaluation, focusing the framework's effectiveness in achieving this goal.

\paragraph{Detecting ambiguities in verbal commands}
Our data collection study (Sec. \ref{sec:data_collection_study}) showed that users often give incomplete or ambiguous instructions.
Our framework's Ambiguity Detection module demonstrated the capability to clarify them, achieving 91.26\% accuracy in determining missing information in our leave-one-user-out cross-validation setting. 
We further tested the module with a leave-one-scenario-out cross-validation, where it achieved an accuracy of 92.93\%.
This result demonstrates its effectiveness in determining whether sufficient information has been gathered to generate a meaningful MR layout.
The ablation study (Sec. \ref{sec:ablation_ambiguity}) also showed that the UI layouts could be better tailored to users' preferences when they provided additional information in response to the clarification questions prompted by our Ambiguity Detection module.

\paragraph{Validating layouts based on verbal instructions}
In our layout comparison study, we assessed whether the Validation module accurately selects the final UI layout according to the user's instructions by comparing its selections with those made by experienced MR participants. 
The results show a 100\% overlap between the layouts identified as best aligning with verbal instructions by both human participants and instances of Validation.
This demonstrates that our framework successfully determines the most useful layout for users from a set of candidates, eliminating the need for them to manually evaluate and select layouts.

\paragraph{Customizing layouts with verbal instructions}
Finally, we analyzed if our end-to-end pipeline could customize UIs based on users' verbal instructions more effectively than the current SOTA, \cbaseline{}, which relies on manual layout selection from the Pareto solution space of a static MOO problem. 
Results from our user study showed 
that participants made fewer adjustments and covered less distance with adjustments before reaching a satisfactory layout with \cours{} compared to \cbaseline{}.



\paragraph{Comparing to manual UI customization}
In our user study, we compared \cours{} with a condition in which users manually adjusted the position of widgets with the controller. 
The NASA-TLX results revealed that while \cours{} led to a higher perceived mental demand, it significantly reduced both physical demand and general effort while achieving comparable performance to \cmanual{}.
Since \cmanual{} allowed participants to adjust the UI layout precisely to their preferences, these findings suggest that \cours{} effectively enabled users to customize their UI layouts. 
We attribute the increased mental workload mainly to the effort of translating thoughts into verbal instructions for the system.
To ease mental demand, future improvements could aim to reduce the number of iterations required to achieve users' goals by leveraging data from their past interactions or preferences.






In summary, study results suggest that our approach of using natural language to customize UI optimization helps users create layouts that better match their preferences while reducing the effort compared to baseline methods.

\section{Limitation \& Future Work}

In the following, we discuss the limitations of our approach and potential directions for future research.

\paragraph{Objective space vs. design space instructions}
We have demonstrated that \projname can generate UI layouts based on users’ instructions. 
However, the framework is currently limited to instructions that relate to the objective space, meaning that instructions are interpreted only in terms of the objective functions currently supported by \projname. 
Commands such as “Place the video player three meters in front of me,” which specify direct positions within the design space (i.e., the parameter space of each virtual widget’s position), are not currently supported.
To overcome this limitation, research should explore new objective functions that can map such design space instructions into the objective space.
Another approach might involve preference optimization focused on learning and adjusting user preferences within the design space \cite{yamamoto2022photographic}. 
A hybrid approach, which infers whether the user's preferences relate to the objective space (e.g., minimizing neck strain) or the design space (e.g., placing elements in specific locations) and adapts the UI accordingly, would also be valuable.

\paragraph{Managing preferences before and after UI optimization}
In our current implementation, \projname optimizes the UI layout based on users’ interaction intentions, selected widgets, and personal preferences. 
These inputs are relatively high-level, typically describing general goals or preferences within the current UI layout. 
Using this information, \projname can reason about the user’s preferences, configure the optimization problem accordingly, and select suitable solutions. 
However, we observed that even after optimization, users often want to make small manual adjustments, such as “move the calendar slightly to the left.” 
These low-level adjustments are difficult to generalize across widgets or scenarios, as they often arise only after users have seen the layout results.
This suggests that not all types of preferences are suitable for expression before optimization, which is also reflected in our study results.
\change{
Accordingly, although \projname aims to interpret and reason about users’ preferences for UI layout before optimization, it can only disambiguate instructions and infer preferences to a limited extent. 
Certain preferences and requirements may only become apparent to users after observing the optimized UI, at which point further refinement is necessary.
}


\paragraph{Preferences elicitation for adaptive UI}
A design feature in \projname that could influence users’ preferences is the clarification conversation. 
The clarification questions are intended to resolve ambiguities in users’ instructions, not simplifying the optimization problem, but to ensure that the optimizer is working towards the user’s true intentions. 
However, when multiple candidate widgets or actions exist, clarification questions could subtly shape or elicit new preferences. 
In our current implementation, we designed the questions to avoid prompting users toward new preferences, but we acknowledge that they may still have encouraged consideration of aspects users had not previously thought about.
For instance, asking whether the user wants to merely watch a widget or also interact with it could lead the user to refine or expand their interaction intention and preferences, which in turn might influence the optimization outcome. 
This highlights an inherent tension faced by adaptive systems: while clarification helps reduce ambiguity, it can also inadvertently shape or bias users’ preferences. 
Future work should investigate methods to navigate this trade-off, ensuring that adaptive UI systems maintain both interpretive clarity and a faithful representation of users’ genuine intentions.

\paragraph{User desired outcomes may not lie on Pareto front}
\change{
In our current implementation, \projname employs six objective functions and three constraints, which are selected from prior work as effective signals for MR UI layout adaptation, but this fixed set may not generalize to all users, tasks, or contexts.
Future research could extend or replace this set to better suit specific use cases. 
}
On the other hand, our optimization module identifies the Pareto front based on the existing objective functions and constraints, and the validation module selects a layout from this front that best aligns with the user’s instructions.
This workflow assumes that the user's preferred interface layout resides on the Pareto front, which may not always hold true. 
The core issue stems from the fact that the objective functions guiding the optimization process are merely mathematical approximations of user preferences. 
While these functions are designed to reflect user priorities, they inevitably fall short of capturing the full complexity of human judgment. 
Consequently, the Pareto front may not encompass all layouts that align with the user's true desires.
To address this limitation, further development of objective functions and constraints that more accurately reflect user preferences could help narrow the gap between optimization outcomes and user expectations.
Meanwhile, our approach supports manual post-optimization adjustments, allowing users to customize interface designs beyond algorithmic solutions.

\begin{figure*}[!h]
  \centering
  \includegraphics[width=\linewidth]{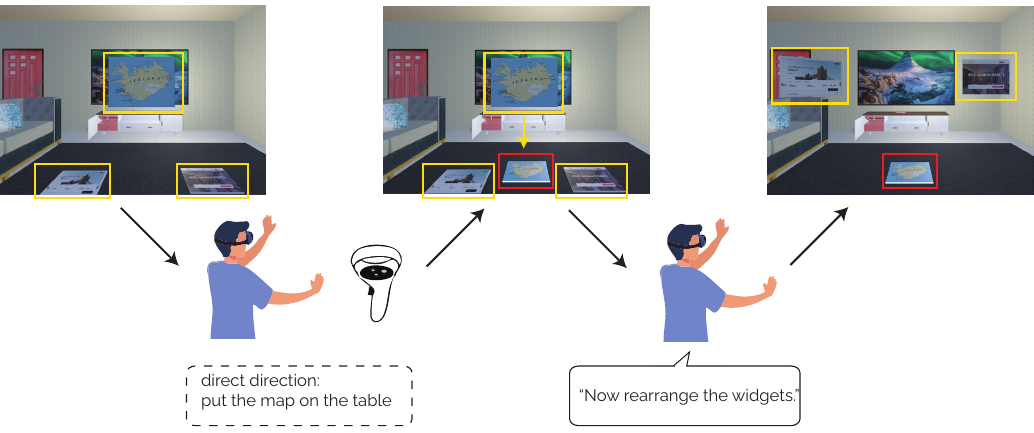}
  \vspace{-3mm}
  \caption{
      Illustration of integrating direct manipulation into \projname. 
      When the user directly changes the position of an element (middle: moving the map from mid-air onto the table) and subsequently provides an additional instruction (e.g., “Now rearrange the widgets.”), \projname incorporates the user’s manipulation into the optimization process. 
      The manipulation result is treated both as the starting point and as a strong constraint that must not be violated. 
      For instance, moving the map away from the TV can be interpreted as a constraint to avoid blocking the TV.
      Based on this constraint, the system then optimizes the placement of the remaining elements (right).
  }
  \label{fig:interaction_optimization}
  \Description{
  The image illustrates an interactive process for optimizing a virtual reality environment. The user, wearing a VR headset and holding a controller, provides instructions to an AI system to arrange objects in a virtual room. In the first panel, two objects—a map and a magazine—are highlighted in yellow boxes on the floor in front of a television. The user gives the command: "put the map on the table". The middle panel shows the result of this command. The map is now on a small table in the center of the room, highlighted by a red box. Another magazine remains on the floor, and a third object has appeared on the table next to the map, all highlighted by red boxes. The original magazine and map on the floor are no longer in their previous positions. In the final panel, the user provides a second command: "Now do not block the TV while keep them visible". The image shows the map and the two magazines arranged horizontally on the table, all in a red box, in a way that does not obstruct the view of the TV.
  }
\end{figure*}

\paragraph{Few-shot learning for Ambiguity Detection}
\change{
Our ambiguity detection module uses few-shot prompting to help the VLM identify unclear aspects of users’ instructions during UI layout optimization, which is in general challenging for VLMs~\cite{ma2025ambigchat}. 
However, prior work shows that in-context learning is highly sensitive to example selection~\cite{zhao2021calibrate, gao2021making, brown2020language}, which can introduce biases that propagate into later configuration decisions.
}
To mitigate this, we included all collected instructions and their annotations as examples, drawing from 29 participants and capturing a diverse range of phrasing styles. 
While this population-level prompt improves general robustness, it may still fail to capture individual user characteristics.
A promising direction is to incorporate continual or user-specific few-shot learning: starting with population examples for initial interactions, and progressively augmenting the context with each user’s own past instructions. 
This approach could make ambiguity detection increasingly personalized over time, potentially improving accuracy and ensuring that subsequent optimization decisions are better aligned with individual preferences.


\paragraph{Evaluation of the Configuration module}
Our framework comprises four components: Ambiguity Detection, Configuration, Optimization, and Validation modules. 
While the optimization module leverages well-established algorithms, we directly evaluated the ambiguity detection and validation modules through meaningful experiments. 
In contrast, evaluating the effectiveness of the Configuration module poses unique challenges. 
Based on user instructions, this module selects virtual widgets, defines the objective terms considered during optimization, and estimates their parameters.

A key challenge in evaluating the Configuration module lies in the absence of ground truth for objective term selection and parameter estimation. 
Since participants may not fully grasp the objective functions employed by the optimization module, they cannot serve as a reliable source of ground truth. 
Consequently, we assess the Configuration module only indirectly, through evaluation of the end-to-end system in our user study. 
The positive results observed, relative to baseline approaches, suggest that the Configuration module effectively captures user intent and environmental context, generating UI configurations that align with users’ expectations.




\paragraph{Alternative approach to problem solving}

Our framework utilizes multiple VLM instances (i.e., agents) within a Chain-of-Thought approach.
However, the capabilities of current language models can be further enhanced by employing more complicated multi-instance structures~\cite{wang2022self, yao2024tree}.
Specifically, we could restructure \projectname as a tree-like hierarchy, where each module represents a node in the tree, generating multiple outputs based on its inputs.

Though computationally more demanding, this approach would enable the framework to self-evaluate, e.g., by comparing outcomes from different branches.
This could result in two key benefits: improved alignment between layouts and user preferences, and reduced need for active user involvement. 
Future work could explore its impact on layout efficiency and accuracy.

\paragraph{Integrating direct manipulation into optimization}

While \projname primarily relies on users’ verbal instructions to guide and optimize the UI layout, it also has the potential to incorporate additional forms of user input to further tailor the interface. 
For example, after customizing the layout based on verbal guidance, the user may interact with the interface to complete the current task, with the flexibility to adjust or reposition virtual elements during interaction. 
When a new round of UI optimization is initiated through verbal input, the framework does not restart from scratch. 
Instead, it incorporates the user’s direct manipulations and manual placements as strong constraints within the optimization process, adjusting the placement of remaining elements accordingly. 
This ensures that both user-driven modifications and contextual considerations are preserved. 
Through this iterative process, the system progressively refines and adapts the interface, fostering a more coherent, personalized, and context-aware optimization workflow, as illustrated in~\autoref{fig:interaction_optimization}.
We acknowledge the incorporation of additional forms of user input as important future work to further customize the automatic optimization process.

\section{Conclusion}

We have presented \projname, a framework that harnesses multi-agentic reasoning to enable personalized UI optimization through natural language instructions.
By dynamically inferring relevant objectives and parameters from verbal instructions, our approach simplifies the complex task of preference-guided multi-objective layout optimization, reducing user effort while improving alignment with individual preferences.

Our evaluation in a MR use case confirmed the framework’s effectiveness. 
The Ambiguity Detection module accurately flagged unclear instructions (91\%), supporting reliable configuration and optimization downstream. 
The Validation module successfully selected final layouts that reflected participants’ preferences, based on the outputs of our Optimization module. 
Together, these components ensured that the generated layouts closely matched user intent, requiring significantly fewer manual adjustments and yielding higher user satisfaction than the baseline method.

In tackling preference-based layout adaptation, \projname addresses limitations of prior approaches by optimizing UIs based on user preferences inferred from natural language instructions. 
By formulating the optimization problem according to these preferences and selecting the Pareto-optimal solution that best aligns with them, our system generates personalized UIs.
This adaptive process reduces both task load and manual customization effort compared to prior approaches, underscoring the potential of \projname for creating individualized adaptive UIs.

\begin{acks}
Yi-Chi Liao was supported by the ETH Zurich Postdoctoral Fellowship Programme.
Zhipeng Li was partially supported by the Swiss National Science Foundation (Grant No. 10004941).
\end{acks}

\bibliographystyle{ACM-Reference-Format}
\bibliography{sample-base}

\clearpage
\appendix
\section{Prompt to support Ambiguity Detection}
\label{app:ambiguity-prompt}

We defined the following prompt to establish the context for the Vision-and-Language Model (VLM). 
The purpose of this prompt is to detect ambiguity in the user's command and, if needed, respond to the user's command with a follow-up question for additional information.


\begin{lstlisting}
To evaluate whether the user's instructions provide sufficient information for customizing Mixed Reality user interfaces, focus on the following criteria:

Primary Widgets: The instructions should clearly identify the virtual widgets that users want to interact with, based on the task at hand. For instance, a user might specify they need a virtual button, slider, or menu.
Interaction Intention: The instructions should describe how users intend to interact with these widgets. This includes whether they want to use direct touch, voice commands, or simply observe the widgets.
User Preference: The instructions should reflect the user's considerations and priorities related to the placement and configuration of virtual widgets. Preferences might include:
Strain and Exertion: Minimizing physical effort or strain.
Field of View: Ensuring widgets are within an optimal viewing area.
Widget Alignment: Aligning widgets in a way that feels natural and intuitive.
Haptic Feedback: Incorporating feedback mechanisms if necessary.
Semantic Connection: Ensuring widgets are logically connected to their functions.
Surface Alignment: Placing widgets on appropriate surfaces.
Situation Context: Adapting widgets to the context in which they will be used.

Categories of Instructions:
Instructions with Sufficient Information: These include clear details on primary widgets, interaction intention, and user preferences. For example, "I need a virtual control panel with sliders for adjusting volume and brightness, which should be interactive through direct touch. The panel should be aligned to the bottom-right of the field of view, considering minimal strain."
Instructions with Missing, Ambiguous, or Irrelevant Information: These lack clarity or completeness in one or more of the key areas. For example, "I want an interface that works well." This instruction does not specify what widgets are needed, how they should be interacted with, or any preferences for their placement or configuration.

Task: Evaluate the following instructions and determine whether they are sufficient or fall into the category of missing, ambiguous, or irrelevant information based on the criteria provided.

The user's instructions are: [user's instructions]
\end{lstlisting}

\noindent
We created the following prompt to provide the VLM with the user's instructions and its corresponding category.

\begin{lstlisting}
instructions with information about the primary virtual widgets, interaction intention and user preference: \ldots
instructions with missing, ambiguous or irrelevant information: \ldots
\end{lstlisting}

\section{Prompt to support Configuration}
\label{app:reasoning-prompt}

Our framework input the following prompt into the VLM alongside the user's instructions for the purpose of selecting a subset of objectives.

\begin{lstlisting}
The current user interface widgets include: [list of widgets]. Based on the user's instructions, please assess whether the following objectives should be considered in the user interface design. For each objective that should be considered, estimate its parameters as described:

Field of View Objective: Determines if virtual widgets should be positioned within the center of the user's field of view. The interval parameter, specified in degrees, represents the user's foveal range.

Arm Angle Objective: Indicates whether virtual widgets should be placed closer to the user's hand position, which is typically lower than the eye position, to facilitate easier interaction.

Neck Angle Objective: Determines if virtual widgets should be positioned near the user's eye level to improve visibility.

Alignment Objective: Assesses whether virtual widgets should be aligned with each other to create a structured user interface. The XTolerance and YTolerance parameters, in meters, define the allowable positional deviations on the X and Y axes.

Physical Anchor Objective: Evaluates whether virtual widgets need to be aligned with a specific physical area or object based on the user's instructions.

Overlay Objective: Determines if virtual widgets should obscure physical objects.

Please provide a detailed analysis based on the description of the user's instruction provided below: [User's instruction description]"}
\end{lstlisting}

\noindent
We created the following prompt for the VLM together with the user's instructions to select virtual widgets and estimate parameters for objective terms.

\begin{lstlisting}
The current user interface includes [number of widgets] virtual widgets. The widgets are: [list of widgets]. The physical areas involved are: [list of areas]. Based on the user's instructions, please determine the following parameters for each virtual widget: 
Interaction Frequency: Represents the user's potential interaction intention with the virtual widget, ranging from 0 (no intention) to 1 (high intention).
Observation Frequency: Indicates the user's potential observation intention with the virtual widget, also ranging from 0 to 1.
Physical Anchor: Specifies whether the virtual widget is anchored to a physical area. The anchor must correspond to one of the provided physical areas. A virtual widget can have no anchor or be assigned to one physical area, and a physical area can be assigned to no more than one virtual widget.
Enabling Setting: Indicates whether the virtual widget will be used based on the user's instructions. If false, the widget will be disabled (i.e., not displayed); if true, the widget will be enabled and visible.

For each physical area, assess the suitability of the area to be occluded or overlaid by virtual widgets based on the user's instructions. If a physical area is unsuitable for occlusion or overlaying, it cannot serve as a physical anchor for any virtual widget. The default overlay suitability should be set to 1 unless specified otherwise.

Please provide the analysis in the same order as the widgets: [list of widgets].

The user's instructions are: [user's instructions]
\end{lstlisting}

\section{Prompt to support Validation}
\label{app:comparison-prompt}

Our framework prompts the VLM with the following query as well as the user's instructions to validate candidate outcomes from the optimizer on the Pareto-front.

\begin{lstlisting}
   
% Compare the following virtual user interface layouts and select the one that best meets the users' requirements. Consider the following factors in your evaluation:

% Ease of Interaction: How easily can users interact with the virtual widgets in each layout? Take into account widget positioning, reachability, and ergonomics.

% Visibility and Observation: How effectively are the virtual widgets positioned within the user's field of view? Consider both primary and peripheral visibility.

% Spatial Arrangement and Alignment: Evaluate the overall structure and alignment of the widgets. Are they organized in a way that is logical and efficient for the user?

% Suitability for Physical Anchoring: Assess how well each layout allows virtual widgets to be anchored to specific physical areas or objects as required by the user.

% Overlay and Occlusion Considerations: Determine if any physical areas are inappropriately occluded or overlaid by virtual widgets, based on the user's instructions.

% Customization and Adaptability: How well does each layout allow for customization or scaling based on user preferences?

% After evaluating these aspects, select the layout that aligns most closely with the users' stated requirements and explain your reasoning.
The user's instructions are: [user's instructions]
\end{lstlisting}

\section{Generated UI layouts in the evaluation}
\label{app: layouts}

\begin{figure*}[ht]
    \centering
    \begin{subfigure}[t]{0.3\textwidth}
        \centering
        \includegraphics[width=\textwidth]{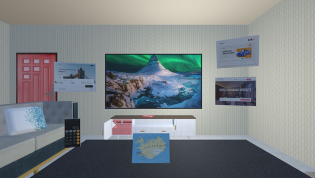}
        \caption{The user's instruction: \textit{"I would like the map to be displayed in front of me, with all other applications positioned around it. They should not obstruct the screen."}}
        \label{fig:subfig1}
    \end{subfigure}
    \hfill
    \begin{subfigure}[t]{0.3\textwidth}
        \centering
        \includegraphics[width=\textwidth]{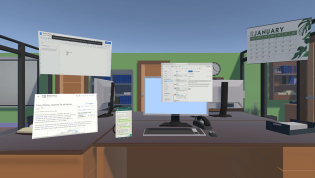}
        \caption{The user's instruction: \textit{"Could you please show everything to me? I want to reply an email first so place it in the center of the monitor."}}
        \label{fig:subfig2}
    \end{subfigure}
    \hfill
    \begin{subfigure}[t]{0.3\textwidth}
        \centering
        \includegraphics[width=\textwidth]{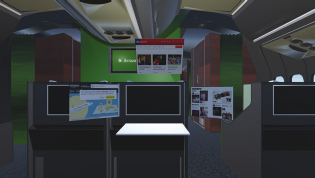}
        \caption{The user's instruction: \textit{"I want to buy a Formula 1 car model on Amazon. Can you also display the sky news? Please do not obstruct my monitor."}}
        \label{fig:subfig3}
    \end{subfigure}
    \vspace{0.5cm}
    \begin{subfigure}[t]{0.3\textwidth}
        \centering
        \includegraphics[width=\textwidth]{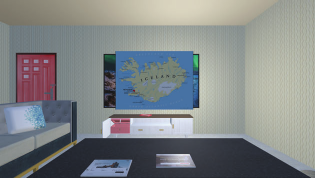}
        \caption{The user's instruction: \textit{"Display the map on the TV, and arrange the flight booking and calculator on the desk so that I can easily interact with them."}}
        \label{fig:subfig4}
    \end{subfigure}
    \hfill
    \begin{subfigure}[t]{0.3\textwidth}
        \centering
        \includegraphics[width=\textwidth]{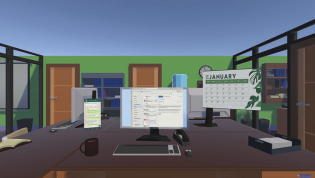}
        \caption{The user's instruction: \textit{"I need to reply to an email. Please put the email on the monitor and position my calendar and messanger in mid-air next to it. Also, be careful about my coffee mug."}}
        \label{fig:subfig5}
    \end{subfigure}
    \hfill
    \begin{subfigure}[t]{0.3\textwidth}
        \centering
        \includegraphics[width=\textwidth]{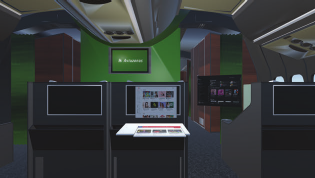}
        \caption{The user's instruction: \textit{"I’d like to watch videos on the TV while scrolling the sports news. I might also play music at the same time, so please position Spotify next to YouTube for easy access."}}
        \label{fig:subfig6}
    \end{subfigure}
    \caption{Instructions and generated layouts in the evaluation study.}
    \Description{The image contains six smaller images arranged in two rows and three columns, each depicting a different computer monitor setup with various user instructions. (a) A monitor with a map application positioned behind an airplane toy, suggesting the toy should not obstruct the screen. (b) A monitor displaying an email client with a note suggesting to place the user’s construction model in front of it without blocking the view. (c) A monitor showing sports footage with a note asking to place Formula 1 model cars without obscuring the screen. (d) A monitor displaying flight booking information with a request to arrange tea on the desk so that it does not obstruct the screen. (e) A monitor showing an email draft and a coffee mug on the desk, accompanied by instructions to send an email and not block the view of the mug. (f) A monitor displaying multiple video streaming services with instructions to position action figures around it without blocking access.}
    \label{fig:2x3grid}
\end{figure*}

\section{Module-Wise Time Cost of \cours{} in the Evaluation}
\label{app: time_cost}


\begin{table}[htb]
    \caption{Time cost (in seconds) of each module in \cours{} during the evaluation study. 
    User input time for instructions is excluded. 
    When the ambiguity detection module asks users to clarify, multiple round ambiguity detection times are averaged. 
    Results are presented as mean (standard deviation).}
    \Description{The image is a table presenting the average time cost in seconds, along with the standard deviation, for different modules within the Auto-Optimization system across three different scenarios: Living room, Office, and Airplane. The time for user input is not included in these figures. The table shows the time costs for four modules: Ambiguity Detection, Configuration, Optimization, and Validation. Ambiguity Detection: This module took the longest in the Office scenario with a mean time of 6.18 seconds and a standard deviation of 2.17. Configuration: The Office scenario had the lowest mean time for this module at 4.35 seconds with a standard deviation of 1.12. Optimization: The Office scenario also had the highest mean time for this module, at 8.46 seconds with a standard deviation of 2.13. The Living room scenario was the next highest at 7.85 seconds (standard deviation of 1.87), while the Airplane scenario was the lowest at 6.38 seconds (standard deviation of 1.29). Validation: All three scenarios had similar mean times for this module, with the Office scenario at 2.34 seconds (standard deviation of 0.39), the Living room at 2.12 seconds (standard deviation of 0.35), and the Airplane at 2.21 seconds (standard deviation of 0.41).}
  \centering
  \small
    \begin{tabular}[width=\columnwidth]{@{}c@{}c@{}c@{}c@{}c@{}}
    \toprule
    \textbf{Scenario~} & \makecell{\textbf{Ambiguity~} \\ \textbf{Detection~}} & \textbf{~Configuration~} & \textbf{~Optimization~} & \textbf{~Validation}\\
    \midrule
    Living room & 5.11(2.34) & 4.54(1.34) & 7.85(1.87) & 2.12(0.35) \\
    Office & 6.18(2.17) & 4.35(1.12) & 8.46(2.13) & 2.34(0.39) \\
    Airplane & 5.68(1.96) & 5.03(1.37) & 6.38(1.29) & 2.21(0.41) \\
    \bottomrule
    \end{tabular}
\end{table}

\end{document}